\documentclass{article}

\usepackage{arxiv}

\usepackage[utf8]{inputenc} 
\usepackage[T1]{fontenc}    

\usepackage{amsmath,amssymb,amsthm,mathtools,bm}
\usepackage{IEEEtrantools}
\usepackage{nicefrac}       
\usepackage{scalerel,stackengine}
\usepackage{float}
\usepackage{graphicx}
\usepackage{booktabs}
\usepackage{siunitx}
\usepackage{doi} 
\usepackage{enumitem}
\usepackage{hyperref}
\usepackage{url}            
\usepackage[nameinlink,noabbrev]{cleveref}
\usepackage{microtype}
\usepackage{siunitx}

\usepackage{algorithm}
\usepackage{algpseudocode}

\usepackage{matlab-prettifier}
\usepackage{listings}

\usepackage{titlesec}
\titleformat{\paragraph}
{\normalfont\normalsize\bfseries}{\theparagraph}{1em}{}
\titlespacing*{\paragraph}
{0pt}{3.25ex plus 1ex minus .2ex}{1.5ex plus .2ex}

\usepackage{cleveref}
\crefname{figure}{Figure}{Figures}
\crefname{table}{Table}{Tables}
\crefname{section}{}{}
\crefname{subsection}{Section}{Sections}
\crefname{subsubsection}{Section}{Sections}
\crefname{equation}{Equation}{Equations}
\crefname{algorithm}{Algorithm}{Algorithms}
\usepackage{etoolbox}
\crefname{section}{}{}

\usepackage[font=scriptsize]{caption}
\hypersetup{
pdftitle={Learning finite viscoelasticity with DAVIS: A supervised framework for generalized standard materials},
pdfsubject={cond-mat.mtrl-sci},
pdfauthor={Simon Wiesheier, Paul Steinmann, Miguel Angel Moreno-Mateos},
pdfkeywords={Data-driven methods, Supervised learning, Constitutive models, Finite strain incompressible viscoelasticity, Automated model discovery, Inverse problems},
}

\hypersetup{
    colorlinks=true,
    linkcolor=blue,     
    filecolor=magenta,  
    urlcolor=black,      
    citecolor=blue,    
}

\newcommand\equalhat{\mathrel{\stackon[1.5pt]{=}{\stretchto{%
    \scalerel*[\widthof{=}]{\wedge}{\rule{1ex}{3ex}}}{0.5ex}}}}

\newcommand{\F}{\mathbf{F}}
\newcommand{\C}{\mathbf{C}}
\newcommand{\Pio}{\mathbf{P}}
\newcommand{\Sio}{\mathbf{S}}
\newcommand{\RR}{\mathbb{R}}

\newcommand{\tr}{\operatorname{tr}}
\newcommand{\cof}{\operatorname{cof}}
\newcommand{\diag}{\operatorname{diag}}
\newcommand{\dev}{\operatorname{dev}}
\newcommand{\sym}{\operatorname{sym}}
\newcommand{\skw}{\operatorname{skw}}
\newcommand{\Lie}{\mathcal{L}}
\newcommand{\dd}{\mathrm{d}}
\newcommand{\sic}{\cdot} 

\newcommand{\lv}[1]{\mathbf{L}_{\mathrm v,#1}}
\newcommand{\dv}[1]{\mathbf{D}_{\mathrm v,#1}}
\newcommand{\wv}[1]{\mathbf{W}_{\mathrm v,#1}}
\newcommand{\de}[1]{\mathbf{d}_{\mathrm e,#1}}
\newcommand{\Mmw}[1]{\mathbf{M}_{\mathrm e,#1}}

\newcommand{\rpotaug}{\widehat{\mathcal{W}}}

\newcommand{\Psifree}{\Psi}  
\newcommand{\Psiaug}{\widehat{\Psi}}
\newcommand{\Psieqb}{\bar{\Psi}^{\mathrm{eq}}}
\newcommand{\Psineqb}{\bar{\Psi}^{\mathrm{neq}}}
\newcommand{\PsineqbI}[1]{\Psineqb_{1,#1}}
\newcommand{\PsineqbII}[1]{\Psineqb_{2,#1}}
\newcommand{\Phic}{\Phi}
\newcommand{\Phimw}[1]{\Phi_{#1}}

\newcommand{\Ioneb}{\bar I_1}
\newcommand{\Itwob}{\bar I_2}
\newcommand{\Ionee}[1]{\bar I_{1,\mathrm e,#1}}
\newcommand{\Itwoe}[1]{\bar I_{2,\mathrm e,#1}}
\newcommand{\Ionetau}[1]{I_{1,\tau,#1}}
\newcommand{\Itwotau}[1]{I_{2,\tau,#1}}
\newcommand{\Jtau}[1]{J_{\tau,#1}}

\newcommand{\Nmw}{N_{\mathrm{mw}}}
\newcommand{\mw}{\alpha}

\newcommand{\Fe}[1]{\mathbf{F}_{\mathrm e,#1}}

\newcommand{\Fv}[1]{\mathbf{F}_{\mathrm v,#1}}
\newcommand{\Fvd}[1]{\dot{\mathbf{F}}_{\mathrm v,#1}}
\newcommand{\Fvinv}[1]{\mathbf{F}_{\mathrm v,#1}^{-1}}
\newcommand{\FvinvT}[1]{\mathbf{F}_{\mathrm v,#1}^{-\top}}

\newcommand{\Ce}[1]{\mathbf{C}_{\mathrm e,#1}}

\newcommand{\Cvinv}[1]{\mathbf{C}_{\mathrm v,#1}^{-1}}
\newcommand{\Cvinvn}[1]{\mathbf{C}_{\mathrm v,#1,n}^{-1}}

\newcommand{\be}[1]{\mathbf{b}_{\mathrm e,#1}}
\newcommand{\betr}[1]{\mathbf{b}_{\mathrm e,#1}^{\mathrm{tr}}}

\newcommand{\Cebar}[1]{\bar{\mathbf{C}}_{\mathrm e,#1}}

\newcommand{\Cedt}[1]{\dot{\mathbf C}_{\mathrm e,#1}}

\newcommand{\tauv}[1]{\boldsymbol{\tau}_{\mathrm{neq},#1}}

\newcommand{\p}{\boldsymbol{\vartheta}}

\theoremstyle{definition}
\newtheorem{remark}{Remark}
\AtEndEnvironment{remark}{\qed}

\begin{document}

\title{Learning finite viscoelasticity with DAVIS: A supervised framework for generalized standard materials}

\author{ 
    \href{https://orcid.org/0009-0008-9625-3446}{\includegraphics[scale=0.06]{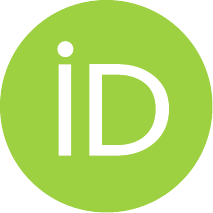}\hspace{1mm}Simon Wiesheier}\thanks{Corresponding author: simon.wiesheier@fau.de} \\
	Institute of Applied Mechanics\\
	Friedrich-Alexander-Universität Erlangen–Nürnberg\\
	91058, Erlangen, Germany\\
    \texttt{simon.wiesheier@fau.de}
    \And
    \href{https://orcid.org/0000-0003-1490-947X}{\includegraphics[scale=0.06]{orcid.pdf}\hspace{1mm}Paul Steinmann} \\
	Institute of Applied Mechanics\\
	Friedrich-Alexander-Universität Erlangen–Nürnberg\\
	91058, Erlangen, Germany\\
	Glasgow Computational Engineering Centre\\
	University of Glasgow\\
	G12 8QQ, United Kingdom\\
	\And
	\href{https://orcid.org/0000-0002-3476-2180}{\includegraphics[scale=0.06]{orcid.pdf}\hspace{1mm}Miguel Angel Moreno-Mateos}\thanks{Corresponding author: miguel.moreno@fau.de} \\
	Institute of Applied Mechanics\\
	Friedrich-Alexander-Universität Erlangen–Nürnberg\\
	91058, Erlangen, Germany\\
	\texttt{miguel.moreno@fau.de} \\
}

\maketitle

\begin{abstract} 
This work revisits the recently proposed data-adaptive viscoelasticity (DAVIS) framework, a spline-based formulation of finite viscoelasticity within the generalized standard materials setting. DAVIS enables a data-driven representation of equilibrium and non-equilibrium constitutive functions while retaining thermodynamic consistency and supporting parameter identification via finite element model updating. The present contribution focuses on improving the robustness and identifiability of non-equilibrium branches in generalized Maxwell-type models. To this end, two extensions of the original formulation are introduced. First, the spline representation is reformulated in terms of curvature-based variables, which is especially convenient to enforce monotonicity and convexity constraints by construction through a smooth parameter mapping. Second, the adaptation of interpolation domains is decoupled from the inner parameter identification by means of a staggered, block-alternating strategy: spline coefficients are optimized for fixed domain endpoints, while the endpoints are updated in an outer loop based on smooth statistics of sampled invariants. This separation alleviates an inherent scaling ambiguity between interpolation domains and spline coefficients that can impair conditioning in viscoelastic inverse problems. The underlying constitutive model remains the finite strain viscoelasticity framework of Reese and Govindjee. The proposed identification strategy is assessed for homogeneous uniaxial loading–unloading tests, which facilitates the study of identifiability and robustness of non-equilibrium branches.
\end{abstract}


\section{Introduction}

Finite deformations, pronounced nonlinearity, and rate dependence are characteristic features of soft material behavior. Their constitutive modeling therefore requires a framework that combines hyperelasticity \cite{steinmann_hyperelastic_2012, ricker_systematic_2023} with viscous dissipation at finite strains \cite{holzapfel_nonlinear_2001}. Among the available approaches, internal-variable formulations based on a multiplicative decomposition of the deformation gradient have proven particularly successful \cite{le_tallec_three-dimensional_1993, lion_physically_1997, reese_theory_1998, lubliner_model_1985}. In this class of models, the finite strain viscoelasticity theory of Reese and Govindjee \cite{reese_theory_1998, govindjee_presentation_1997} has emerged as a widely adopted constitutive backbone. Despite differences in formulation, many of these models can be shown to be closely related when expressed in a common configuration \cite{gouhier_comparison_2024}.

While these formulations provide a sound mechanical basis, the identification of suitable constitutive functions remains a central challenge. Classical approaches require the analyst to commit a priori to specific functional forms, often leading to a trade-off between expressiveness, robustness, and interpretability. This limitation has motivated increasing interest in data-driven constitutive modeling \cite{kirchdoerfer_data-driven_2016}, where constitutive relations are inferred directly from data. For hyperelasticity, a wide range of approaches has emerged, including neural networks \cite{fuhg_review_2025,kalina_automated_2022,klein_polyconvex_2022,klein_neural_2025,linden_neural_2023,thakolkaran_can_2025,abdolazizi_constitutive_2025}, Gaussian processes \cite{frankel_tensor_2019}, symbolic regression \cite{abdusalamov_automatic_2023}, neural ordinary differential equations \cite{tac_data-driven_2023}, EUCLID \cite{flaschel_automated_2023}, material fingerprinting \cite{flaschel_unsupervised_2026}, and spline-based methods, included our framework based on spline-based strain-energy density functions \cite{sussman_model_2009,dal_data-driven_2023, tikenogullari_data-driven_2023, wiesheier_versatile_2024, moreno-mateos_biaxial_2025, wiesheier_data-adaptive_2026}.

For viscoelasticity, however, the task is more demanding: both equilibrium and non-equilibrium contributions must be identified together with a thermodynamically admissible evolution law. In this context, the framework of \textit{generalized standard materials} (GSM), introduced by \cite{halphen_sur_1975} and rooted in the fundamental results developed, amongst others, by \cite{ziegler_thermodynamik_1957, ziegler_attempt_1958}, provides a natural foundation. GSM describes material behavior through two scalar functions, the \textit{free energy} and a \textit{dissipation potential}. The latter can equivalently be expressed in dual form, parameterized in terms of thermodynamic forces, with both formulations linked via the Legendre--Fenchel transformation. It was shown by \cite{kumar_two-potential_2016} that internal-variable formulations of finite strain viscoelasticity based on multiplicative decompositions can be recast within this framework, thereby highlighting its generality and thermodynamic consistency. This structure is particularly attractive for data-driven modeling, as it allows physical constraints to be enforced by construction.

Early data-driven approaches within this framework include variational Onsager neural networks \cite{yu_onsagernet_2021, huang_variational_2022}, which parameterize both energy and dissipation potentials. These studies, however, were limited to one-dimensional settings. More recent GSM-based approaches remained restricted to geometrically linear theory and therefore cannot describe soft material behavior \cite{flaschel_automated_2023-1,rosenkranz_viscoelasticty_2024}.

For finite deformations, most current data-driven viscoelasticity models instead rely on the multiplicative decomposition. Representative examples include neural ODE formulations \cite{tac_data-driven_2023} and recent neural-network-based models by Kalina et al. \cite{kalina_physics-augmented_2026}, and Holthusen et al. \cite{holthusen_complement_2026, holthusen_generalized_2026}. These approaches enforce key physical principles, such as incompressibility of the total deformation (\(J=1\)), isochoric inelastic flow (\(J_{\mathrm v}=1\)), and thermodynamic admissibility within a learning framework. Other directions include formulations based on external-state variables \cite{upadhyay_physics-informed_2024,upadhyay_visco-hyperelastic_2020}, where thermodynamic admissibility is enforced via inequality constraints during training, as well as approaches building on generalized Prony series \cite{abdolazizi_viscoelastic_2024}.

In contrast, to the best of the authors' knowledge, \textit{the DAVIS framework} (Data-Adaptive VIScoelasticity) \textit{is the first approach that combines a data-driven representation of the two GSM potentials with constitutive functions based on the multiplicative decomposition for finite viscoelasticity}. It was also the first to achieve this through an unsupervised calibration of the potentials within the GSM formalism \cite{wiesheier_data-adaptive_2026-1}.\footnote{The original DAVIS framework leveraged finite element model updating \cite{mahnken_unified_1996} together with reaction forces and heterogeneous full-field displacement measurements \cite{sutton_determination_1983, avril_overview_2008} to enrich the invariant space that is sampled.} Similar to neural-network-based approaches, DAVIS enforces incompressibility of the total deformation and isochoric inelastic flow, employs implicit time integration, and satisfies polyconvexity requirements.

The original DAVIS formulation demonstrated that data-adaptive spline representations can be successfully embedded into a thermodynamically consistent finite strain viscoelasticity framework. Importantly, unlike black-box models (including non-interpretable neural-network models), \textit{the direct data-driven identification of free-energy and dissipation-potential functions as spline interpolants endows DAVIS with inherent mechanical and physical interpretability}. When extending the approach to richer generalized Maxwell structures, however, challenges arise; they are primarily related to identifiability \cite{hartmann_identifiability_2018} and numerical conditioning. In particular, the simultaneous optimization of spline coefficients and interpolation-domain endpoints introduces a scaling ambiguity between constitutive response and parameter domain. While manageable for moderate model complexity, this coupling becomes increasingly restrictive as the number of Maxwell branches grows. In addition, the value-based spline representation requires explicit inequality constraints to enforce polyconvexity \cite{hartmann_polyconvexity_2003, schroder_invariant_2003}, which increases the complexity of the inverse problem.

\begin{figure}
    \centering
    \includegraphics[width=0.93\linewidth]{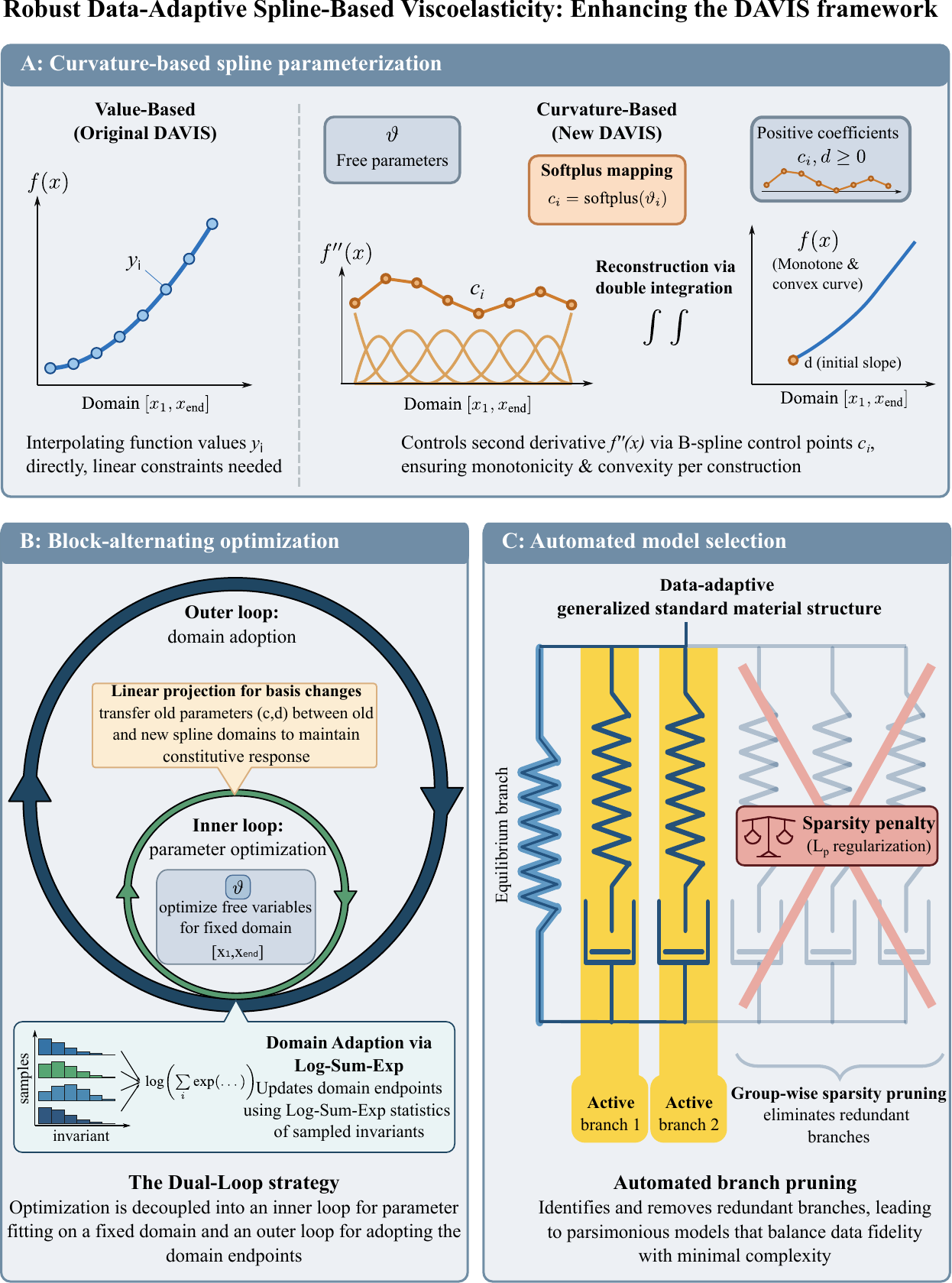}
    \caption{\textbf{Schematic overview of the proposed extensions to the DAVIS framework.} 
    (a) Curvature-based spline representation of the constitutive functions, in which convexity and monotonicity are enforced by construction through non-negative curvature coefficients and boundary slopes. 
    (b) Block-alternating identification strategy, separating the optimization of spline parameters from the adaptation of the interpolation domains. (c) Group-wise sparsity promotion across Maxwell branches, enabling automatic identification of a parsimonious viscoelastic model by selectively activating relevant branches.}
    \label{fig:schematics}
\end{figure}

The present work builds on DAVIS and introduces several enhancements that improve the robustness, scalability, and interpretability of the identification procedure, i.e.,
\begin{itemize}
    \item First, the spline representation is reformulated in terms of curvature-like variables and boundary slopes, ensuring polyconvexity by construction and eliminating the need for explicit inequality constraints. 
    \item Second, the adaptation of interpolation domains is decoupled from the parameter identification through a block-alternating strategy based on smooth statistics of sampled invariants.
    \item Third, a group-wise sparsity regularization is introduced to promote the automatic selection of relevant Maxwell branches, cf.~\cref{fig:schematics}.
\end{itemize}

In contrast to the original DAVIS work, which focused on full finite element model updating, the present study deliberately restricts attention to homogeneous uniaxial loading--unloading experiments. In machine learning terminology, this corresponds to supervised training based on stress--strain data. This provides a controlled setting to systematically investigate identifiability, assess the effect of the proposed reformulation, and study the automatic selection of non-equilibrium branches.

\section{Finite strain incompressible viscoelasticity}
\label{sec:finite_visco}

The present formulation is based on the finite strain viscoelasticity theory of Reese and Govindjee~\cite{reese_theory_1998}, specialized to incompressible isotropic materials.  It is recast here in the language of generalized standard materials~\cite{halphen_sur_1975}: the constitutive model is fully characterized by a \emph{free energy} $\Psifree$ governing the recoverable response and branch-wise \emph{dissipation potentials} $\{\Phimw{\mw}\}$ governing the irreversible evolution. These two ingredients are sufficient to derive all constitutive equations from a single variational principle at the material point. The separation into $(\Psifree,\Phimw{\mw})$ is particularly valuable in the present data-adaptive setting, as it identifies precisely the constitutive functions to be represented by splines.

\subsection{Constitutive setting}
The present formulation employs the following notation for the principal kinematic, stress, and thermodynamic quantities:
\begin{table}[H]
\centering
\caption{Principal notation used in the finite-strain viscoelastic formulation.}
\label{tab:notation_visco}
\renewcommand{\arraystretch}{1.3}
\begin{tabular}{ll}
\hline
Symbol & Meaning / property \\
\hline
$\F,\; J=\det(\F)=1$
& deformation gradient and incompressibility \\
$\C=\F^\top \sic \F,\; \bar{\C}=J^{-2/3}\C$
& right Cauchy--Green tensor and its isochoric part \\
$\Ioneb,\; \Itwob$
& invariants of the isochoric total strain tensor \\
$\F=\Fe{\mw}\sic \Fv{\mw}$
& multiplicative decomposition for branch $\mw$ \\
$\Fv{\mw},\; J^{\mathrm v(\mw)}=\det(\Fv{\mw})=1$
& viscous distortion for branch $\mw$ \\
$\Fe{\mw},\; J^{\mathrm e(\mw)}=\det(\Fe{\mw})=1$
& elastic distortion for branch $\mw$ \\
$\Ce{\mw}=\Fe{\mw}^{\top}\sic \Fe{\mw},\;
\be{\mw}=\Fe{\mw}\sic \Fe{\mw}^{\top}$
& elastic right and left Cauchy--Green tensors \\
$\Cebar{\mw}=\bigl[\det(\Ce{\mw})\bigr]^{-1/3}\Ce{\mw}$
& isochoric elastic right Cauchy--Green tensor \\
$\Ionee{\mw},\; \Itwoe{\mw}$
& invariants of the isochoric elastic branch tensor \\
$\lv{\mw},\;
\dv{\mw}=\sym(\lv{\mw}),\;
\wv{\mw}=\skw(\lv{\mw})$
& viscous velocity gradient and its symmetric/skew parts \\
$\tauv{\mw},\; \dev(\tauv{\mw})=\tauv{\mw}$
& purely deviatoric non-equilibrium Kirchhoff stress \\
$\Ionetau{\mw},\; \Itwotau{\mw},\; \Jtau{\mw}$
& invariants of the non-equilibrium Kirchhoff stress \\
$\de{\mw},\; \tr(\de{\mw})=0$
& purely deviatoric viscous deformation-type rate \\
$\Psifree,\; \Psieqb,\; \Psineqb_{\mw}$
& free energy, equilibrium and non-equilibrium contributions \\
$\Pi_{\mw},\; \Phimw{\mw}$
& primal and dual dissipation potentials \\
$p,\; \Psiaug=\Psifree-p[J-1]$
& hydrostatic pressure and augmented free energy \\
$\rpotaug,\; \mathcal{D}\ge0$
& augmented rate potential and reduced dissipation \\
$\Mmw{\mw},\; \Sio,\; \Pio=\F\sic \Sio$
& Mandel, Piola--Kirchhoff and Piola stresses \\
\hline
\end{tabular}
\end{table}

\subsubsection{Kinematics and invariants}

We consider a generalized Maxwell model with one equilibrium branch and $\Nmw$ non-equilibrium branches arranged in parallel. All branches share the total deformation, whereas branch-wise internal variables describe viscous effects. For each non-equilibrium branch \(\mw=1,\dots,\Nmw\), the deformation gradient is multiplicatively decomposed as
\begin{equation}
\F=\Fe{\mw}\sic\Fv{\mw}.
\label{eq:mult_decomp}
\end{equation}

This decomposition introduces an intermediate (stress-free) configuration associated with the viscous mapping \(\Fv{\mw}\), separating reversible (elastic) and dissipative (viscous) mechanisms. Equivalently, the internal state may be described by the inverse viscous tensor $\Cvinv{\mw}=\Fvinv{\mw}\sic\FvinvT{\mw}$, which induces the elastic right and left Cauchy--Green tensors
\begin{equation}
\Ce{\mw}
=
\FvinvT{\mw}\sic\C\sic\Fvinv{\mw},
\qquad
\be{\mw}
=
\F\sic\Cvinv{\mw}\sic\F^\top,
\label{eq:elastic_CG}
\end{equation}
\noindent with $\C=\F^\top\sic\F$.

The branch-wise viscous velocity gradient and its symmetric and skew-symmetric parts are
\begin{equation}
  \lv{\mw} := \Fvd{\mw}\sic\Fvinv{\mw},
  \qquad
  \dv{\mw} := \sym(\lv{\mw}),
  \qquad
  \wv{\mw} := \skw(\lv{\mw}).
  \label{eq:lv_split}
\end{equation}

Differentiating $\Ce{\mw} = \FvinvT{\mw}\sic\C\sic\Fvinv{\mw}$ with respect to time yields the kinematic identity
\begin{equation}
  \Cedt{\mw}
  = \FvinvT{\mw}\sic\dot{\C}\sic\Fvinv{\mw}
    - \lv{\mw}^{\top}\sic\Ce{\mw}
    - \Ce{\mw}\sic\lv{\mw}
  \label{eq:Ce_dot}
\end{equation}
which links the rate of the elastic strain measure to the observable rate $\dot{\C}$ and the viscous velocity gradient $\lv{\mw}$.

To enforce incompressibility and isolate isochoric deformation, we apply the Flory split~\cite{flory_thermodynamic_1961}
\begin{equation}
  \bar{\C} = J^{-2/3}\C,
  \qquad
  \Cebar{\mw} = \bigl[\det(\Ce{\mw})\bigr]^{-1/3}\Ce{\mw},
  \qquad
  J = \det(\F).
  \label{eq:flory}
\end{equation}

Assuming isotropy, the constitutive functions depend on these tensors only through their principal invariants. For the isochoric total tensor, we introduce
\begin{equation}
  \Ioneb = \tr(\bar{\C}),
  \qquad
  \Itwob = \bigl[\tr(\cof\bar{\C})\bigr]^{3/2} - 3\sqrt{3},
  \label{eq:inv_bulk}
\end{equation}
and for the isochoric elastic branch tensors
\begin{equation}
  \Ionee{\mw} = \tr(\Cebar{\mw}),
  \qquad
  \Itwoe{\mw} = \bigl[\tr(\cof\Cebar{\mw})\bigr]^{3/2} - 3\sqrt{3},
  \label{eq:inv_branch}
\end{equation}
where $\cof(\cdot)$ denotes the cofactor. The transformed second invariant follows~\cite{hartmann_polyconvexity_2003, schroder_invariant_2003} and is particularly convenient for constructing polyconvex constitutive functions.

\subsubsection{Rate-type variational potential}
\label{sec:variational}

The constitutive equations are derived from a rate-type variational principle at the material point, in the spirit of \cite{ortiz_variational_1999, fancello_variational_2006, fancello_variational_2008, mosler_variationally_2010, miehe_homogenization_2002}. In this framework, the model is governed by a scalar rate potential whose stationarity conditions simultaneously yield the internal variable evolution and identify the stress. The observable rate $\dot{\C}$ enters as a \emph{prescribed external input} and is not subject to variation at the material-point level; the internal variable rates are the free quantities.
 
\medskip
\noindent\textit{Augmented free energy and rate potential.}\quad
The incompressibility constraint $J = 1$ is incorporated into the variational structure by introducing the Lagrange multiplier $p$ (hydrostatic pressure) and defining the \emph{augmented free energy}
\begin{equation}
  \Psiaug\bigl(\bar{\C},\{\Cebar{\mw}\},p\bigr)
  :=
  \Psifree\bigl(\bar{\C},\{\Cebar{\mw}\}\bigr)
  - p\,[J-1],
  \label{eq:psi_aug}
\end{equation}
where $\Psifree$ is the free energy defined in \cref{eq:free_energy} below, and $p$ is not a constitutive parameter but a reactive field determined by the incompressibility condition. In the present setting, it is determined by an additional problem-specific constraint, e.g., a vanishing transverse stress component under homogeneous uniaxial loading. We collect the constitutive state in
\begin{equation}
  \mathsf{z}
  := \bigl(\C,\{\Fv{\mw}\}_{\mw=1}^{\Nmw},p\bigr),
\end{equation}
and define the \emph{augmented rate potential} at fixed state
$\mathsf{z}$ as%
\footnote{At fixed state, the mapping $\Fvd{\mw}\mapsto\lv{\mw}=\Fvd{\mw} \cdot \Fvinv{\mw}$ is linear and invertible, so $\lv{\mw}$ is used as the branch rate variable in place of $\Fvd{\mw}$.}
\begin{equation}
  \rpotaug\!\bigl(\dot{\C},\{\lv{\mw}\},\dot{p};\mathsf{z}\bigr)
  :=
  \dot{\Psiaug} + \sum_{\mw=1}^{\Nmw}\Pi_{\mw}\!\bigl(\dv{\mw}\bigr).
  \label{eq:rate_potential}
\end{equation}

Here $\Pi_{\mw}$ is the branch-wise dissipation potential specified in \cref{sec:constitutive_model} below.
 
\medskip
\noindent\textit{Reduced rate potential.}\quad
To make the stationarity conditions explicit, we need the material time derivative of $\Psiaug$. At this stage, we keep the free energy $\Psifree$ generic, assuming only that it depends on the total strain measure $\C$ and on the branch-wise elastic strain measures $\{\Ce{\mw}\}$; the specific functional forms are introduced later in \cref{sec:constitutive_model}. Differentiating~\cref{eq:psi_aug} with respect to time and applying the chain rule gives
\begin{equation}
  \dot{\Psiaug}
  =
  -[J-1]\dot{p}
  - p\,\dot{J}
  + \frac{\partial\Psifree}{\partial\C}:\dot{\C}
  + \sum_{\mw=1}^{\Nmw}
    \frac{\partial\Psifree}{\partial\Ce{\mw}}:\Cedt{\mw}.
  \label{eq:psi_aug_chainrule}
\end{equation}

The key step is to substitute the kinematic identity~\cref{eq:Ce_dot} into the last sum. Each branch term splits immediately into two physically distinct contributions:
\begin{equation}
  \frac{\partial\Psifree}{\partial\Ce{\mw}}:\Cedt{\mw}
  =
  \underbrace{%
    \Bigl[
      \Fvinv{\mw}\sic
      \frac{\partial\Psifree}{\partial\Ce{\mw}}
      \sic\FvinvT{\mw}
    \Bigr]:\dot{\C}%
  }_{\text{stress contribution}}
  -
  \underbrace{%
    \frac{\partial\Psifree}{\partial\Ce{\mw}}
    :\bigl[
      \lv{\mw}^{\top}\sic\Ce{\mw}
      +\Ce{\mw}\sic\lv{\mw}
    \bigr]%
  }_{\text{dissipation contribution}}.
  \label{eq:branch_split}
\end{equation}

The first group is proportional to $\dot{\C}$ and contributes to the stress, while the second group is proportional to $\lv{\mw}$ and drives the viscous dissipation. It can be simplified further: since $\Ce{\mw}$ is symmetric and $\Psifree$ is isotropic in $\Ce{\mw}$, the tensor $\partial\Psifree/\partial\Ce{\mw}$ is symmetric and coaxial with $\Ce{\mw}$. We therefore introduce the symmetric Mandel-type branch stress~\cite{platen_nonlinear_2024}
\begin{equation}
  \Mmw{\mw}
  :=
  2\,\Ce{\mw}\sic\frac{\partial\Psifree}{\partial\Ce{\mw}}
  =
  2\,\frac{\partial\Psifree}{\partial\Ce{\mw}}\sic\Ce{\mw},
  \label{eq:mandel}
\end{equation}
in terms of which the dissipation contribution reads
\begin{equation}
  \frac{\partial\Psifree}{\partial\Ce{\mw}}
  :\bigl[\lv{\mw}^{\top}\sic\Ce{\mw}+\Ce{\mw}\sic\lv{\mw}\bigr]
  = \Mmw{\mw}:\dv{\mw},
  \label{eq:md_identity}
\end{equation}
where only the symmetric part $\dv{\mw}=\sym(\lv{\mw})$ survives because $\Mmw{\mw}$ is symmetric; the skew part $\wv{\mw}$ drops out entirely. Inserting~\cref{eq:branch_split,eq:md_identity} into~\cref{eq:psi_aug_chainrule} and using $\dot{J}=\tfrac{1}{2}J\,\C^{-1}:\dot{\C}$ to absorb the $-p\dot{J}$ term into the $\dot{\C}$-coefficient gives
\begin{equation}
  \dot{\Psiaug}
  =
  -[J-1]\dot{p}
  + \tfrac{1}{2}\,\Sio:\dot{\C}
  - \sum_{\mw=1}^{\Nmw}\Mmw{\mw}:\dv{\mw},
  \label{eq:psi_dot_final}
\end{equation}
where $\Sio$ is the Piola--Kirchhoff stress, defined as the total coefficient of $\tfrac{1}{2}\dot{\C}$:
\begin{equation}
  \Sio
  :=
  -Jp\,\C^{-1}
  + 2\,\frac{\partial\Psifree}{\partial\C}
  + 2\sum_{\mw=1}^{\Nmw}
    \Fvinv{\mw}\sic
    \frac{\partial\Psifree}{\partial\Ce{\mw}}
    \sic\FvinvT{\mw}.
  \label{eq:PK2}
\end{equation}

The three terms represent, respectively, the reactive pressure contribution, the stress from the total strain energy, and the non-equilibrium branch stresses pulled back from the intermediate configuration to the reference configuration. The explicit form of $\Sio$ in terms of the specific free energy contributions
$\Psieqb$ and $\{\Psineqb_{\mw}\}$ follows once those functions are introduced in \cref{sec:constitutive_model}. The Piola stress is given by $\Pio = \F\sic\Sio$.

Substituting~\cref{eq:psi_dot_final} into~\cref{eq:rate_potential} yields the reduced augmented rate potential
\begin{equation}
  \rpotaug
  =
  -[J-1]\dot{p}
  + \tfrac{1}{2}\,\Sio:\dot{\C}
  + \sum_{\mw=1}^{\Nmw}
    \Bigl[\Pi_{\mw}(\dv{\mw}) - \Mmw{\mw}:\dv{\mw}\Bigr].
  \label{eq:rate_potential_reduced}
\end{equation}

\medskip
\noindent\textit{Stationarity conditions.}\quad
The constitutive equations follow from stationarity of $\rpotaug$ with respect to the internal rates $\{\lv{\mw}\}$\footnote{Although $\Pi_{\mw}$ depends on $\dv{\mw}=\sym(\lv{\mw})$, the variation is taken with respect to $\lv{\mw}$. Since $\delta\dv{\mw}=\sym(\delta\lv{\mw})$ and $\partial\Pi_{\mw}/\partial\dv{\mw}$ is symmetric, one has $\frac{\partial\Pi_{\mw}}{\partial\dv{\mw}}:\sym(\delta\lv{\mw}) =\frac{\partial\Pi_{\mw}}{\partial\dv{\mw}}:\delta\lv{\mw}$, so the skew part of $\delta\lv{\mw}$ does not contribute.} and the multiplier rate $\dot{p}$, and from differentiation with respect to the rate $\dot{\C}$:
\begin{subequations}\label{eq:stationarity_conditions}
\begin{IEEEeqnarray}{rClCll}
\text{Incompressibility:}
&\quad&
\delta_{\dot{p}}\rpotaug
&=&
-[J-1]\,\delta\dot{p}
=0
\qquad
\forall\,\delta\dot{p}
&
\nonumber\\
&&
&\Longrightarrow&
J=1 ,
&
\label{eq:stat_p}
\\[8pt]
\text{Evolution equations:}
&\quad&
\delta_{\lv{\mw}}\rpotaug
&=&
\Bigl[
\frac{\partial\Pi_{\mw}}{\partial\dv{\mw}}
-
\Mmw{\mw}
\Bigr]
:
\delta\lv{\mw}
=0
\qquad
\forall\,\delta\lv{\mw},
&\quad
\mw=1,\dots,\Nmw
\nonumber\\
&&
&\Longrightarrow&
\Mmw{\mw}
=
\frac{\partial\Pi_{\mw}}{\partial\dv{\mw}},
&\quad
\mw=1,\dots,\Nmw ,
\IEEEeqnarraynumspace
\label{eq:stat_lv}
\\[8pt]
\text{Stress relation:}
&\quad&
\frac{\partial\rpotaug}{\partial\dot{\C}}
&=&
\tfrac{1}{2}\,\Sio .
&
\label{eq:stress_id}
\end{IEEEeqnarray}
\end{subequations}

~\cref{eq:stat_lv} is the flow rule for each non-equilibrium branch: the Mandel stress $\Mmw{\mw}$ is thermodynamically conjugate to the viscous rate $\dv{\mw}$ and is balanced by the dissipative resistance $\partial\Pi_{\mw}/\partial\dv{\mw}$. ~\cref{eq:stat_p} enforces the incompressibility constraint. The stress~\cref{eq:stress_id} is not obtained by variation but by differentiation, cf. \cite{fancello_variational_2006}.%

\begin{remark}
The variation with respect to $\dot{\C}$ would correspond to varying the applied load, which has no meaning at the material-point level. The comparison of the \textit{material} variational principle utilized in this work with a \emph{global} (or structural) variational principle clarifies this. At the structural level one varies the displacement field $\mathbf{u}$ (or equivalently
$\dot{\mathbf{u}}$) over the body $\mathcal{B}$, which yields the weak form of linear momentum balance,
$\int_{\mathcal{B}} \Pio : \nabla\delta\dot{\mathbf{u}}\,\dd V = \int_{\partial\mathcal{B}} \bar{\mathbf{t}}\cdot\delta\dot{\mathbf{u}}\,\dd A$, or equivalently $\operatorname{Div}\Pio = \mathbf{0}$ in strong form. There, $\dot{\mathbf{u}}$ is the unknown, and its variation drives the equilibrium equations. However, at the material point, $\dot{\mathbf{u}}$ (or $\dot{\C}$) is an \emph{input}, not an unknown. The stress is therefore identified by differentiation with respect to $\dot{\C}$, rather than by variation as in a global variational principle.
\end{remark}
 
The reduced dissipation takes the form
\begin{equation}
  \mathcal{D}
  = \sum_{\mw=1}^{\Nmw}\Mmw{\mw}:\dv{\mw} \ge 0,
  \label{eq:dissipation}
\end{equation}
confirming that $(\Mmw{\mw},\dv{\mw})$ is the natural force--rate pair in the intermediate configuration.
 
\begin{remark}[Equivalent spatial representation]
  \label{rem:spatial}
  For the specific constitutive forms introduced in \cref{sec:constitutive_model}, it is convenient to rewrite the branch equations in the spatial configuration. Introducing the non-equilibrium Kirchhoff stress \footnote{In the present incompressible setting, the Kirchhoff stress coincides with the Cauchy stress.}
  \begin{equation}
    \tauv{\mw}
    = 2\,\Fe{\mw}\sic
      \frac{\partial\Psineqb_{\mw}}{\partial\Ce{\mw}}
      \sic\Fe{\mw}^{\top},
    \label{eq:tau_neq}
  \end{equation}
  and the spatial viscous deformation-type rate \footnote{$\mathcal{L}(\bullet)$ denotes the Lie derivative, representing the objective rate of the elastic left Cauchy--Green tensor.}
  \begin{equation}
    \de{\mw}
    := -\tfrac{1}{2}\,\Lie(\be{\mw})\sic\be{\mw}^{-1},
    \label{eq:de_spatial}
  \end{equation}
  one verifies the equivalence $\Mmw{\mw}:\dv{\mw} = \tauv{\mw}:\de{\mw}$, so that $(\tauv{\mw},\de{\mw})$ is an alternative conjugate pair for the same dissipative mechanism. The spatial form of the evolution ~\cref{eq:stat_lv} then reads $\tauv{\mw} = \partial\Pi_{\mw}/\partial\de{\mw}$, which is the form used in the incremental update of \cref{sec:incremental_update}.
\end{remark}

\subsubsection{Constitutive model}
\label{sec:constitutive_model}

Having established the variational structure, we now specify the two constitutive ingredients of the generalized standard material: the free energy $\Psifree$ and the dissipation potentials $\{\Phimw{\mw}\}$.
Together, these functions completely define the model; all stress and evolution expressions follow from the framework of \cref{sec:variational}.
 
\medskip
\noindent\textit{\textbf{Free energy:}}\quad
The free energy is additively decomposed into an equilibrium contribution and $\Nmw$ non-equilibrium branch contributions,
\begin{equation}
  \Psifree\bigl(\bar{\C},\{\Cebar{\mw}\}\bigr)
  =
  \Psieqb(\bar{\C})
  +
  \sum_{\mw=1}^{\Nmw}\Psineqb_{\mw}(\Cebar{\mw}),
  \label{eq:free_energy}
\end{equation}
where the dependence on the isochoric tensors $\bar{\C}$ and $\Cebar{\mw}$ defined in~\cref{eq:flory} guarantees objectivity and invariance with respect to arbitrary rotations of the intermediate
configuration. Using the invariants in~\cref{eq:inv_bulk,eq:inv_branch}, each contribution is further split as
\begin{align}
  \Psieqb(\bar{\C})
  &= \Psieqb_1(\Ioneb) + \Psieqb_2(\Itwob),
  \label{eq:psi_eq}
  \\[2pt]
  \Psineqb_{\mw}(\Cebar{\mw})
  &= \PsineqbI{\mw}(\Ionee{\mw}) + \PsineqbII{\mw}(\Itwoe{\mw}).
  \label{eq:psi_neq}
\end{align}

The use of invariant-based representations ensures objectivity and material symmetry. The scalar functions $\Psieqb_1$, $\Psieqb_2$, $\PsineqbI{\mw}$, $\PsineqbII{\mw}$ are the constitutive functions to be identified from data; in the present work, they are represented by splines.
 
\medskip
\noindent\textit{\textbf{Dissipation potentials:}}\quad
We now specify the branch-wise dissipation potential $\Pi_{\mw}$ introduced in the rate potential~\cref{eq:rate_potential}. In the material description, $\Pi_{\mw}$ is a convex function of the viscous rate $\dv{\mw}$; in the equivalent spatial description of \cref{rem:spatial} it is expressed as a convex \emph{primal dissipation potential} $\varphi_{\mw}(\de{\mw})$, with $\Pi_{\mw}(\dv{\mw}) = \varphi_{\mw}(\de{\mw})$ under the push-forward equivalence $\Mmw{\mw}:\dv{\mw} = \tauv{\mw}:\de{\mw}$. Equivalently, one may work with the \textit{dual dissipation potential} obtained by the Legendre--Fenchel transform,
\begin{equation}
  \Phimw{\mw}(\tauv{\mw})
  =
  \sup_{\de{\mw}}
  \Bigl\{
    \tauv{\mw}:\de{\mw} - \varphi_{\mw}(\de{\mw})
  \Bigr\},
  \label{eq:LF}
\end{equation}
from which the complementary viscous evolution reads
\begin{equation}
  \de{\mw} = \frac{\partial\Phimw{\mw}}{\partial\tauv{\mw}}.
  \label{eq:flow_rule_dual}
\end{equation}

In the present work, we adopt the \emph{dual dissipation potential} because it allows the dissipative response to be parameterized directly in terms of stress invariants \footnote{Both the primal and dual dissipation potentials provide equivalent thermodynamic descriptions of the inelastic process. However, the dual formulation is often preferred because it yields an explicit evolution law and is typically more convenient for numerical implementation \cite{leuschner_potential-based_2015}.}. Thermodynamic admissibility requires $\Phimw{\mw}$ to be convex, non-negative, and normalized so that $\Phimw{\mw}(\mathbf{0}) = 0$.
 
For isotropic materials in the incompressible setting, volumetric dissipation is excluded, and the potential depends only on the deviatoric part of $\tauv{\mw}$~\cite{stewart_large_2024}. Introducing the
stress invariants
\begin{equation}
  \Ionetau{\mw} = \tr(\tauv{\mw}),
  \qquad
  \Itwotau{\mw} = \tr(\cof\tauv{\mw}),
  \qquad
  \Jtau{\mw} = \Ionetau{\mw}^{2} - 3\,\Itwotau{\mw},
  \label{eq:stress_invs}
\end{equation}
we take
\begin{equation}
  \Phimw{\mw} = \Phimw{\mw}(\Jtau{\mw}).
  \label{eq:phi_form}
\end{equation}

The invariant $\Jtau{\mw}$ admits the equivalent form $\Jtau{\mw} = \tfrac{3}{2}\dev\tauv{\mw}:\dev\tauv{\mw}$, confirming that it captures purely deviatoric information and vanishes for
volumetric states. Furthermore, $\Jtau{\mw}$ is convex in $\tauv{\mw}$~\cite{tac_data-driven_2023}, so any convex scalar function $\Phimw{\mw}(\Jtau{\mw})$ defines a thermodynamically admissible dual
potential. As with the free energy, $\Phimw{\mw}$ is represented by a spline in the present work.
 
\begin{remark}
    More generally, the dual dissipation potential may be expressed as an isotropic scalar function of the deviatoric part of $\tauv{\mw}$, which admits two independent invariants since $\det{\tauv{\mw}} = 1$. The single-invariant form~\cref{eq:phi_form} is adopted here for parsimony; numerical results did not indicate the need for a richer representation.
\end{remark}
 
\begin{remark}
  For a compressible extension of the framework including volumetric dissipation, the reader is referred to~\cite{li_large-deformation_2022,liu_large_2025}.
\end{remark}

\subsection{Implicit incremental branch update}
\label{sec:incremental_update}

The temporal evolution of each non-equilibrium branch is integrated by means of an exponential mapping algorithm. This choice preserves symmetry and positive definiteness of \(\be{\mw}\) throughout the update. As shown below, the present formulation yields a purely isochoric viscous flow, so that the viscous deformation remains unimodular, consistent with standard assumptions for incompressible materials.

\paragraph{Trial state and exponential update}
Given the internal state at time \(t_n\) and the deformation \(\F_{n+1}\) at the end of the current increment, one first forms the trial elastic state
\begin{equation}
\betr{\mw} = \F_{n+1} \sic \Cvinvn{\mw} \sic \F_{n+1}^{\top}.
\end{equation}

This trial state corresponds to freezing the viscous evolution during the increment. The updated elastic left Cauchy--Green tensor is obtained by correcting the trial state through the viscous flow via the exponential map,
\begin{equation}
\be{\mw,n+1}
=
\exp\!\big(-2\Delta t\,\de{\mw}\big) \sic \betr{\mw},
\label{eq:exp_update}
\end{equation}
where $\de{\mw}$ is given by the dual flow rule~\cref{eq:flow_rule_dual}.

\paragraph{Spectral local problem}
By isotropy, $\be{\mw}$, $\betr{\mw}$ and $\tauv{\mw}$ are coaxial and the update~\cref{eq:exp_update} can be performed in their common principal basis. Denoting principal quantities by superscripts $(\cdot)^A$, $A = 1,2,3$, the viscous evolution reduces component-wise to $(d_{\mathrm{e},\mw})^A = \partial\Phimw{\mw}/\partial\tau_{\mw}^A$, and the residual equations read
\begin{equation}
  r_{\mw}^A
  =
  \varepsilon_{\mathrm{e},\mw}^A
  - (\varepsilon_{\mathrm{e},\mw}^{\mathrm{tr}})^A
  + \Delta t\,(d_{\mathrm{e},\mw})^A
  = 0,
  \qquad A = 1,2,3,
  \label{eq:local_residual}
\end{equation}
where $\varepsilon_{\mathrm{e},\mw}^A = \ln(\lambda_{\mathrm{e},\mw}^A)$ are the logarithmic principal elastic strains from the spectral decomposition $\be{\mw} = \sum_{A}(\lambda_{\mathrm{e},\mw}^A)^2\,
\mathbf{n}_A\otimes\mathbf{n}_A$. The residual is implicit because $\tau_{\mw}^A$ depends on the updated elastic state, and the resulting three-dimensional nonlinear system is solved by a local Newton method at each material point.

\paragraph{Isochoric viscous flow}
Since $\Phimw{\mw}$ depends only on the deviatoric invariant $\Jtau{\mw}$, the dual flow rule~\cref{eq:flow_rule_dual} yields a traceless rate, $\operatorname{tr}(\de{\mw}) = 0$. Consequently,
\begin{equation}
  \det\bigl(\exp(-2\Delta t\,\de{\mw})\bigr) = 1,
\end{equation}
and the exponential update preserves $\det(\be{\mw})$:
\begin{equation}
  \det(\be{\mw,n+1}) = \det(\betr{\mw}).
\end{equation}

For incompressible loading starting from a unimodular state, this implies $\det(\be{\mw}) = 1$ throughout, which via $\be{\mw} = \F\sic\Cvinv{\mw}\sic\F^{\top}$ is equivalent to
\begin{equation}
  \det(\Fv{\mw}) = 1,
\end{equation}
i.e., the viscous deformation remains unimodular, consistent with the incompressible framework.

\section{Robust data-adaptive spline-based viscoelasticity}
\label{sec:data_adaptive_finite_visco}

The constitutive framework of Section~\cref{sec:finite_visco} reduces the material description to a finite collection of scalar functions within the GSM formalism, namely the equilibrium and non-equilibrium energy contributions and the branchwise dissipation potentials:
\begin{itemize}[noitemsep]
    \item Equilibrium contributions:
    \[
    \Psieqb_1(\Ioneb), \qquad \Psieqb_2(\Itwob),
    \]
    \item Non-equilibrium energy contributions for each branch $\mw=1,\dots,\Nmw$:
    \[
    \Psineqb_{1,\mw}(\Ionee{\mw}), \qquad \Psineqb_{2,\mw}(\Itwoe{\mw}),
    \]
    \item Dual dissipation potentials, also for each branch $\mw=1,\dots,\Nmw$:
    \[
    \Phimw{\mw}(\Jtau{\mw}).
    \]
\end{itemize}

\textit{In the DAVIS framework \cite{wiesheier_data-adaptive_2026-1}, these potentials are not prescribed analytically, but represented in a data-adaptive manner by spline functions defined on invariant spaces}. The present section first recalls this basic philosophy and then introduces the modifications that form the core of this work.

In the original DAVIS formulation for finite strain viscoelasticity, the constitutive functions were represented by B-spline interpolants \cite{d_boor_practical_1978, butterfield_computation_1976} whose function values at interpolation sites served as optimization variables. The interpolation domains were adapted together with the spline parameters, and calibration was performed in an unsupervised setting by finite element model updating using reaction forces and full-field displacement data. While this strategy proved effective for moderately sized models, it revealed robustness issues for richer generalized Maxwell structures. In particular, the simultaneous optimization of spline values and domain endpoints introduces an ambiguity between the scaling of the constitutive functions and the scaling of the interpolation domains. Moreover, the value-based representation requires explicit inequality constraints to enforce polyconvexity.

The present contribution revisits DAVIS from this perspective. We retain the constitutive backbone of Section~\cref{sec:finite_visco} and the central idea of representing the scalar constitutive ingredients by splines on invariant spaces, but reformulate the identification problem in three respects:
\begin{itemize}
\item The spline representation is recast in terms of curvature-like quantities and boundary slopes, so that the required structural properties are enforced by construction.
\item The adaptation of the interpolation domains is decoupled from the inner parameter fit by a block-alternating strategy based on invariant samples generated by the current forward solution.
\item The resulting robust parameterization enables a sparsity-promoting extension for automatic branch selection.
\end{itemize}

In contrast to the original DAVIS setting, we restrict attention here to homogeneous uniaxial loading--unloading experiments, i.e., supervised training. The simplicity of the calibration data as stress-strain data provides a clean benchmark for studying the identifiability of the viscoelastic branches.

\subsection{Curvature-based spline representation}

Let
\begin{equation}
    f : [x_1,x_{\mathrm{end}}] \to \RR
\end{equation}
denote a generic scalar constitutive function. In the present context, \(f\) stands for one of the functions
\[
\Psieqb_1,\quad \Psieqb_2,\quad
\Psineqb_{1,\mw},\quad \Psineqb_{2,\mw},\quad
\Phic_{\mw}.
\]

We consider the spline space \(\mathcal{S}_p(\Xi)\) of degree \(p\) on the interval \([x_1,x_{\mathrm{end}}]\), defined by the knot vector
\begin{equation}
    \Xi = \{\xi_1,\dots,\xi_m\},
    \qquad
    \xi_1 = \dots = \xi_{p+1} = x_1 \le \dots \le \xi_{m-p} = \dots = \xi_m = x_{\mathrm{end}}.
\end{equation}
Let \(\{N_i\}_{i=1}^{n_b}\) denote the associated B-spline basis. Then any \(f \in \mathcal{S}_p(\Xi)\) admits the representation
\begin{equation}
    f(x) = \sum_{i=1}^{n_b} a_i N_i(x).
\end{equation}

In the original DAVIS formulation, the constitutive function is parameterized by its values at interpolation sites,
\begin{equation}
    f(x_i) = y_i,
    \qquad i=1,\dots,n,
\end{equation}
which induces the representation
\begin{equation}
    f(x) = \sum_{i=1}^{n} y_i \tilde N_i(x)
\end{equation}
in an interpolation basis \(\{\tilde N_i\}\). While this provides direct control over function values, structural properties such as convexity and monotonicity are not intrinsic and must be enforced by additional inequality constraints on \(\mathbf y\).

\medskip

To overcome this limitation, we adopt a curvature-based parameterization. Instead of representing \(f\) itself, we prescribe its second derivative via free variables $\vartheta_i$,
\begin{equation}
    f''(x) = \sum_{i=1}^{n_c} c_i N_i(x),
    \qquad c_i = \operatorname{softplus}(\vartheta_i),
\end{equation}
and reconstruct the function by integration,
\begin{equation}
\begin{aligned}
    f'(x) &= d + \int_{x_1}^{x} f''(s)\,\dd s,
    \qquad d = \operatorname{softplus}(\vartheta_0), \\
    f(x)  &= f(x_1) + \int_{x_1}^{x} f'(s)\,\dd s.
\end{aligned}
\end{equation}

The additive constant \(f(x_1)\) is fixed by constitutive normalization.

\medskip

This representation yields the following structural properties:
\begin{itemize}
\item \(N_i(x)\ge 0\), \(c_i\ge 0 \Rightarrow f''(x)\ge 0\), hence \(f\) is convex.
\item \(d\ge 0 \Rightarrow f'(x)\ge 0\), hence \(f\) is monotone.
\end{itemize}

Thus, convexity and monotonicity are enforced by construction, without explicit inequality constraints.

For each scalar constitutive function \(f\), we introduce the parameter vector
\begin{equation}
    \p^{(f)} = (\vartheta_0,\vartheta_1,\dots,\vartheta_{n_c})^\top \in \RR^{n_c+1},
\end{equation}
where \(\vartheta_0\) defines the boundary slope and the remaining entries determine the curvature. The global parameter vector is obtained by concatenation,
\begin{equation}
    \p =
    \bigl(
        \p^{(\Psieqb_1)},\;
        \p^{(\Psieqb_2)},\;
        \{\p^{(\Psineqb_{1,\mw})}\}_{\mw=1}^{\Nmw},\;
        \{\p^{(\Psineqb_{2,\mw})}\}_{\mw=1}^{\Nmw},\;
        \{\p^{(\Phic_{\mw})}\}_{\mw=1}^{\Nmw}
    \bigr).
\end{equation}

Compared to the value-based formulation, admissibility constraints are embedded directly into the representation, and the optimization variables are unconstrained up to simple numerical bounds preventing overflow of the softplus mapping.

\subsection{Adaptive spline domains: Block-alternation decoupling}

In the original DAVIS framework, the right endpoint \(x_{\mathrm{end}}\) of each spline domain was optimized together with the spline parameters utilizing Kernel-Density-Estimation \cite{silverman_density_2018}. For convex increasing constitutive functions, however, this leads to an inherent scaling ambiguity. 

Indeed, consider a linear function $f(x)=ax$ on $[0,x_{\mathrm{end}}]$. A rescaling of the domain $x \mapsto \tilde x = \gamma x$ can be compensated by scaling the function values, so that the slope over the rescaled domain remains unchanged. Consequently, different combinations of domain endpoints and spline coefficients may produce nearly identical responses over the invariant range activated by the data. This results in flat directions in the objective function and deteriorates identifiability.

\begin{remark}
For the equilibrium energy, the admissible invariant range is directly determined by the calibration data, so that the spline domain can be fixed a priori, cf.~\cite{wiesheier_data-adaptive_2026,moreno-mateos_learning_2026}. In contrast, the non-equilibrium contributions depend on the internal evolution and are therefore only implicitly defined through the local constitutive update. This makes the effective invariant range solution-dependent and motivates the adaptive domain strategy.

For arguments beyond the spline domain, we employ a controlled extrapolation by keeping the curvature constant, i.e., \(f''(x)=f''(x_{\mathrm{end}})\) for \(x>x_{\mathrm{end}}\). This yields a linear continuation of \(f'(x)\) and avoids uncontrolled growth of the constitutive response.
\end{remark}

To remove this coupling, we decouple parameter identification and domain adaptation. Instead of solving a monolithic optimization problem in \((\p,x_{\mathrm{end}})\), we employ a block-alternating strategy. Given a current domain endpoint \(x_{\mathrm{end}}^{(k)}\), we first solve the inner problem
\begin{equation}
    \p^{(k+1)}
    =
    \arg\min_{\p}
    \mathcal{J}\bigl(\p,x_{\mathrm{end}}^{(k)}\bigr),
\end{equation}
where \(\mathcal{J}\) denotes the loss functional introduced later. The resulting forward solution yields a set of sampled invariants
\begin{equation}
    \mathcal{X}^{(k+1)}
    =
    \{x_q^{(k+1)}\}_{q=1}^{N_{\mathrm{samp}}},
\end{equation}
collected over all evaluation points and, in the general setting, all datasets. These samples characterize the activated region of the invariant space and are used to update the spline domain in an outer step,
\begin{equation}
    x_{\mathrm{end}}^{(k+1)}
    =
    \mathcal{U}\!\left(x_{\mathrm{end}}^{(k)},\mathcal{X}^{(k+1)}\right).
\end{equation}

To define the update operator \(\mathcal{U}\), we employ a smooth upper-tail statistic based on the log-sum-exp functional,
\begin{equation}
    x_{\mathrm{act}}
    =
    \frac{1}{\alpha}
    \log\!\left(
        \frac{1}{N_{\mathrm{samp}}}
        \sum_{q=1}^{N_{\mathrm{samp}}}
        \exp(\alpha x_q)
    \right),
\end{equation}
where \(\alpha>0\) is a smoothing parameter. For large \(\alpha\), this quantity approaches the sample maximum, whereas for moderate values it remains differentiable and less sensitive to isolated extreme values. The updated endpoint is then taken as
\begin{equation}
    x_{\mathrm{end}}^{(k+1)}
    =
    [1-\eta]\,x_{\mathrm{end}}^{(k)} + \eta\,x_{\mathrm{act}},
\end{equation}
where \(\eta\in(0,1]\) is a relaxation parameter. This choice allows gradual expansion or contraction of the spline domain over the outer iterations.

After updating the endpoint \(x_{\mathrm{end}}\), the spline domain \([x_1,x_{\mathrm{end}}]\) is linearly rescaled, which induces a change of the underlying spline basis. As a consequence, the current representation of the constitutive function must be transferred to the updated domain.

The derivative admits the representation
\begin{equation}
    f'(x)
    =
    d + \sum_{i=1}^{n_c} c_i
    \int_{x_1}^{x} N_i(s)\,\dd s.
\end{equation}

To transfer the representation, we preserve the slope over the sampled invariants \(\mathcal{X}=\{x_q\}_{q=1}^{N_{\mathrm{samp}}}\),
\begin{equation}
    f'_{\mathrm{new}}(x_q) \approx f'_{\mathrm{old}}(x_q),
    \qquad q=1,\dots,N_{\mathrm{samp}}.
\end{equation}

This leads to the non-negative least-squares problem: find \(\mathbf q = (d,\mathbf c)\ge 0\) such that
\begin{equation}
    \mathbf q
    =
    \arg\min_{\mathbf q \ge 0}
    \sum_{q=1}^{N_{\mathrm{samp}}}
    \left[
        d + \sum_{i=1}^{n_c} c_i
        \int_{x_1}^{x_q} N_i(s)\,\dd s
        - f'_{\mathrm{old}}(x_q)
    \right]^2.
\end{equation}

\begin{remark}
The projection is carried out in the physical parameters $(d,\mathbf c)$ to retain a linear least-squares problem. Performing it in the free variables would result in a non-linear problem without benefit.
\end{remark}

The resulting algorithm alternates between parameter optimization on fixed domains and domain updates informed by the current invariant samples. By separating spline shape from domain scaling, this strategy removes a major source of non-identifiability and improves the conditioning of the inverse problem. The algorithmic structure is summarized in~\cref{alg:block_alt}. 

\begin{algorithm}[htbp!]
\caption{Block-alternating spline domain adaptation}
\label{alg:block_alt}
\begin{algorithmic}[1]
\State Initialize \(\p^{(0)}\) and spline-domain endpoints \(x_{\mathrm{end}}^{(0)}\)
\State Set \(k \gets 0\)

\Repeat
    \State \textbf{Inner step:}
    \[
        \p^{(k+1)}
        =
        \arg\min_{\p}
        \mathcal{J}\bigl(\p,x_{\mathrm{end}}^{(k)}\bigr)
    \]

    \State Perform forward simulations with \(\p^{(k+1)}\)
    \State Collect sampled invariants \(\mathcal{X}^{(k+1)}\)
    \State Compute smooth upper-tail statistics \(x_{\mathrm{act}}^{(k+1)}\)
    \State Update spline-domain endpoints
    \[
        x_{\mathrm{end}}^{(k+1)}
        =
        \mathcal{U}\bigl(x_{\mathrm{end}}^{(k)},x_{\mathrm{act}}^{(k+1)}\bigr)
    \]

    \If{\(x_{\mathrm{end}}^{(k+1)} \neq x_{\mathrm{end}}^{(k)}\)}
        \State Convert \(\p^{(k+1)}\) to physical parameters \((c,d)\)
        \State Project \((c,d)\) onto the updated spline basis by non-negative least squares
        \State Map the projected physical parameters back to \(\p^{(k+1)}\)
    \EndIf

    \State \(k \gets k+1\)
\Until{the relative decrease of \(\mathcal{J}\) and the change of \(x_{\mathrm{end}}\) are both below prescribed tolerances}

\State \textbf{Final refinement:}
\[
    \min_{\p}
    \mathcal{J}\bigl(\p,x_{\mathrm{end}}^{(k)}\bigr)
\]
with fixed spline domains
\end{algorithmic}
\end{algorithm}

\subsection{Loss function: Identification from homogeneous uniaxial loading--unloading data}

In contrast to the original DAVIS formulation, which relied on finite element model updating with reaction forces and full-field displacement data, we focus here on homogeneous uniaxial loading--unloading experiments, i.e., supervised training. This setting provides a transparent benchmark for studying the identifiability of the branchwise constitutive functions and the effect of the proposed reformulation.

Let
\begin{equation}
    \mathcal{D} = \{1,\dots,N_{\mathrm{set}}\}
\end{equation}
denote the set of experimental datasets. For each \(s\in\mathcal{D}\), a prescribed stretch history \(\lambda^{(s)}(t)\) and measured Piola stresses
\(
P^{\mathrm{exp},(s)}(t_i^{(s)})
\)
at discrete times \(\{t_i^{(s)}\}_{i=1}^{N_s}\) are given. Under homogeneous uniaxial loading in direction \(\boldsymbol e_1\), the deformation gradient is
\begin{equation}
    \F \equalhat \diag(\lambda,\lambda_\perp,\lambda_\perp),
    \qquad
    \lambda \lambda_\perp^2 = 1.
\end{equation}

For a given parameter vector \(\p\), the constitutive update of Section~\cref{sec:finite_visco} yields the model prediction
\begin{equation}
    P^{\mathrm{mod},(s)}(t_i^{(s)};\p)
    = P_{11}.
\end{equation}

The identification problem is formulated as
\begin{equation}
    \min_{\p}
    \mathcal{J}_{\mathrm{data}}(\p),
\end{equation}
with the dataset-wise normalized least-squares functional
\begin{equation}
    \mathcal{J}_{\mathrm{data}}(\p)
    =
    \frac{1}{2}
    \sum_{s\in\mathcal{D}}
    \frac{
        \sum_{i=1}^{N_s}
        \left[
            P^{\mathrm{mod},(s)}(t_i^{(s)};\p)
            -
            P^{\mathrm{exp},(s)}(t_i^{(s)})
        \right]^2
    }{
        \sum_{i=1}^{N_s}
        \bigl[P^{\mathrm{exp},(s)}(t_i^{(s)})\bigr]^2
    }.
\end{equation}

This normalization ensures comparable contributions of all datasets. The optimization is performed over \(\p\), while the spline domains are updated in an outer loop.

\begin{remark}
The optimization problems were solved using a gradient-based least-squares method (\textsc{Matlab} \texttt{lsqnonlin}). No numerical instabilities such as NaNs or Infs were observed during the identification process. This contrasts with recent neural-network-based viscoelastic models, where such issues may arise due to highly nonlinear parameterizations and poorly controlled initializations. In the present approach, the curvature-based spline representation admits simple and physically meaningful initial guesses, leading to a well-conditioned optimization problem.
\end{remark}

Compared to the original DAVIS formulation, the present approach differs in two essential respects. First, the constitutive functions are represented by the curvature-based parameterization rather than by spline values, so that convexity and monotonicity are enforced intrinsically. Second, the interpolation domains are not part of the inner optimization variables, but are adapted from invariant samples generated by the current forward solve.

\subsection{Sparsity-promoting branch selection}

Once the inverse problem is robust with respect to the spline representation and domain adaptation, one may further address the question of model complexity. In particular, the number of non-equilibrium Maxwell branches need not be fixed a priori if redundant branches can be pruned automatically during identification. To this end, we augment the data misfit by a group-wise sparsity term acting at the level of whole Maxwell branches,
\begin{equation}
    \mathcal{J}_{\mathrm{tot}}(\p)
    =
    \mathcal{J}_{\mathrm{data}}(\p)
    +
    \lambda_{\mathrm{sparse}}
    \sum_{\mw=1}^{\Nmw}
    \mathcal{G}_{\mw}(\p),
\end{equation}
where \(\lambda_{\mathrm{sparse}}\ge 0\) is a regularization parameter.

For each branch \(\mw\), the penalty acts on the physical parameters of the associated constitutive functions
\[
    \Psineqb_{1,\mw},\qquad
    \Psineqb_{2,\mw},\qquad
    \Phic_{\mw},
\]
that is, on the curvature coefficients and boundary slopes defining these functions. Collecting these physical parameters into a vector
\begin{equation}
    \mathbf q_{\mw}
    =
    \bigl(
        d_{1,\mw},\,\mathbf c_{1,\mw},\,
        d_{2,\mw},\,\mathbf c_{2,\mw},\,
        d_{\Phi,\mw},\,\mathbf c_{\Phi,\mw}
    \bigr),
\end{equation}
we define the normalized group penalty as a smooth $\ell^p$-type quasi-norm,
\begin{equation}
    \mathcal{G}_{\mw}
    =
    \left[
        \sum_{i=1}^{n_q}
        [q_{\mw,i}+\delta]^p
    \right]^{1/p},
    \qquad 0<p\le 1,
\end{equation}
where $\delta>0$ is a small smoothing parameter and $n_q$ denotes the number of parameters in $\mathbf q_{\mw}$. This choice promotes sparsity at the level of entire branches rather than individual spline coefficients.

\begin{remark}
The smoothing parameter $\delta>0$ ensures differentiability of the penalty and thus enables the use of gradient-based optimization methods. In the limit $\delta \to 0$, the formulation approaches a non-smooth $\ell^p$ quasi-norm, which promotes stronger sparsity but is not compatible with the present gradient-based solver.
\end{remark}

After identification, branch activity is assessed from the derivatives of the identified constitutive functions evaluated over the sampled invariant values $\{x_q\}_{q=1}^{N_{\mathrm{samp}}}$. For branch $\mw$, we define
\begin{align}
    A_{1,\mw}^2
    &= \frac{1}{N_{\mathrm{samp}}}
    \sum_{q=1}^{N_{\mathrm{samp}}}
    \left[
        \frac{\mathrm d}{\mathrm d x}
        \Psineqb_{1,\mw}(x_q)
    \right]^2,\\
    A_{2,\mw}^2
    &= \frac{1}{N_{\mathrm{samp}}}
    \sum_{q=1}^{N_{\mathrm{samp}}}
    \left[
        \frac{\mathrm d}{\mathrm d x}
        \Psineqb_{2,\mw}(x_q)
    \right]^2,\\
    A_{\Phi,\mw}^2
    &= \frac{1}{N_{\mathrm{samp}}}
    \sum_{q=1}^{N_{\mathrm{samp}}}
    \left[
        \frac{\mathrm d}{\mathrm d x}
        \Phic_{\mw}(x_q)
    \right]^2,
\end{align}
and set
\begin{equation}
    A_{\mw}
    =
    \left[
        A_{1,\mw}^2 + A_{2,\mw}^2 + A_{\Phi,\mw}^2
    \right]^{1/2}.
\end{equation}
over the activated invariant range. A branch is classified as inactive if
\begin{equation}
    \frac{A_{\mw}}{\max_k A_k} < \varepsilon_{\mathrm{act}},
\end{equation}
with prescribed threshold \(\varepsilon_{\mathrm{act}}\).

\begin{remark}
If the derivatives of the non-equilibrium constitutive functions vanish over the activated invariant range, the corresponding branch does not contribute to the stress response. The quantity $A_{\mw}$ therefore provides a natural measure of branch activity.
\end{remark}

\section{Numerical examples and discussion}
\label{sec:results}

We consider a set of homogeneous uniaxial loading--unloading experiments at different stretch rates. Each dataset \(s \in \mathcal{D}\) consists of a prescribed stretch history \(\lambda^{(s)}(t)\) and Piola stress \(P^{\mathrm{exp},(s)}(t)\). Specifically, we employ experimental data for VHB Tape 4910 reported in~\cite{hossain_experimental_2012}, which comprise loading rates of \(\dot{\lambda} = \SI{0.01}{\per\second}\), \(\dot{\lambda} = \SI{0.03}{\per\second}\) and \(\dot{\lambda} = \SI{0.05}{\per\second}\) for different maximum prescribed stretches.\footnote{The experimental data were taken from the repository \url{https://github.com/ConstitutiveANN/vCANN}.}

The identification problem is formulated as described in Section~\cref{sec:data_adaptive_finite_visco}, where all datasets are incorporated simultaneously through the normalized least-squares functional. The calibration is performed using the block-alternating strategy introduced above.

For later discussion, we denote by
\begin{equation}
    \mathcal{I}(f) := [x_1^{(f)},x_{\mathrm{end}}^{(f)}]
\end{equation}
the final spline domain associated with a scalar constitutive function \(f\). In the present formulation, \(x_1^{(f)}\) is fixed by the constitutive normalization, whereas \(x_{\mathrm{end}}^{(f)}\) is adapted during the outer iterations.

In the following, we analyze the robustness, interpretability, and scalability of the proposed formulation.

\subsection{Original DAVIS formulation exhibits robustness limitations}

As a baseline, we consider the original DAVIS formulation based on value parameterization of spline functions combined with a joint optimization of spline values and domain endpoints.

We deliberately choose a minimal, yet representative model consisting of a single Maxwell element and \(n=5\) interpolation values per spline for each constitutive function
\[
\Psineqb_{1,1}(\Ionee{1}), 
\qquad 
\Psineqb_{2,1}(\Itwoe{1}), 
\qquad 
\Phic_{1}(\Jtau{1}).
\]

This setup is not intended to yield an optimal physical model but rather to isolate the behavior of the optimization procedure. A robust formulation should converge to the same solution for a wide range of initializations.

To assess robustness, we perform a \textit{deterministic multi-start study}. The initial splines are constructed as piecewise linear functions with zero reference value and constant slope increments. The slope increments of the three constitutive contributions
\[
\Psineqb_{1,1}, \qquad \Psineqb_{2,1}, \qquad \Phic_{1}
\]
are scaled independently using
\[
(s_{I_1},\, s_{I_2},\, s_{\Phi}) \in \{0.3,\,1,\,3\}^3,
\]
resulting in \(27\) deterministic initializations probing both magnitude and relative shape variations. For each initialization, the constrained nonlinear least-squares problem is solved and the final objective value is recorded.

\begin{figure}
    \centering
    \includegraphics[width=0.98\linewidth]{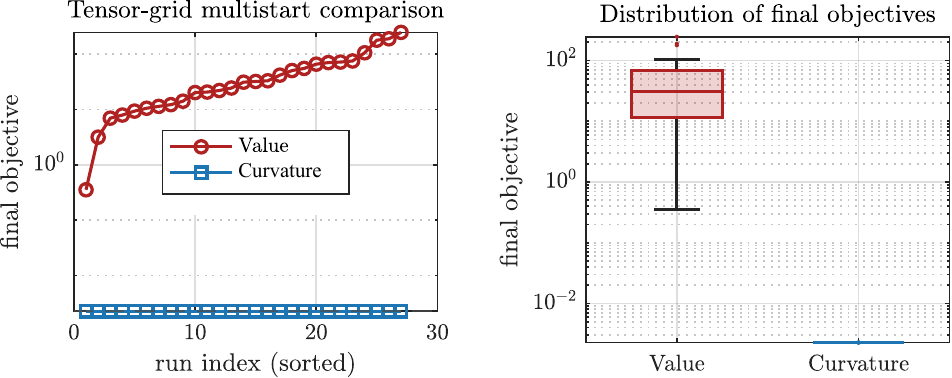}
    \caption{\textbf{Multi-start robustness study.} Final objective values obtained from a deterministic tensor grid of initializations for the original value-based DAVIS formulation and for the proposed method. The value-based formulation exhibits a pronounced spread of final objective values, whereas enhanced DAVIS consistently converges to the same solution.}
    \label{fig:multistart}
\end{figure}

The results in~\cref{fig:multistart} show a pronounced sensitivity of the value-based formulation to the initial guess, with a wide spread of final objective values. None of the runs converges to the solution obtained by the proposed method. A closer inspection reveals that the optimization typically does not terminate based on first-order optimality conditions. Instead, the step size is progressively reduced until stagnation occurs. This behavior indicates the presence of nearly flat directions in the objective landscape.

Flat directions in the objective landscape arise from the simultaneous optimization of spline values and domain endpoints, which introduces a scaling ambiguity between function magnitude and domain size. As a result, different parameter combinations yield almost identical model responses, leading to ill-conditioning and unreliable convergence. In addition, the enforcement of convexity and monotonicity by linear inequality constraints further complicates the optimization landscape.

\subsection{Enhanced DAVIS yields robust and well-conditioned optimization}

We now consider the proposed formulation with curvature-based spline representation and adaptive spline domains. The block-alternating scheme is configured with seven outer iterations and 20 inner optimization steps per outer iteration. During each inner loop, the spline domains are fixed. After completion of the outer iterations, a final refinement phase of 50 inner iterations is performed. Furthermore, a multi-start study analogous to the previous subsection shows that all initializations converge to the same solution, indicating that the proposed reformulation removes the dominant source of ill-conditioning observed in the original DAVIS approach.

\begin{figure}
    \centering
    \includegraphics[width=0.85\linewidth]{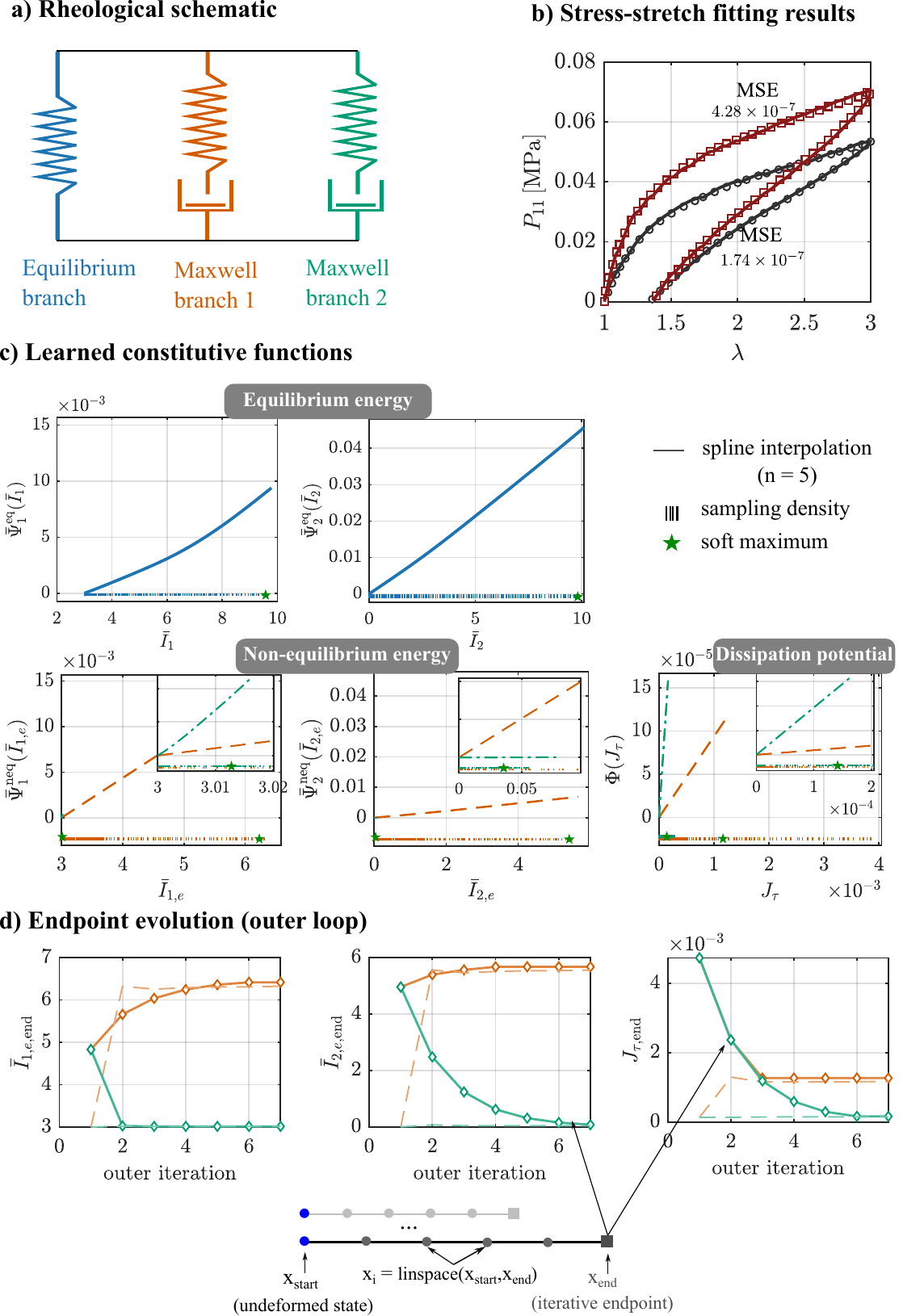}
    \caption{\textbf{Calibration results for a model with two Maxwell elements and \(n=5\) spline coefficients per scalar constitutive function.} (a) Rheological schematic. (b) Stress--stretch curves for both datasets together with the identified model response. (c) Identified constitutive spline functions; rug plots indicate the sampled invariant values obtained from the forward simulation with the calibrated model, and the soft-maximum target values are highlighted. (d) Evolution of the domain endpoints \(x_{\mathrm{end}}^{(f)}\) during the seven outer iterations; dashed lines denote the log-sum-exp targets and solid lines the actual endpoint values. Colors introduced in panel (a) consistently identify the branches of the rheological model and their associated contributions throughout the figure.}
    \label{fig:Nmw_3_n_5}
\end{figure}

The resulting calibrated model is shown in~\cref{fig:Nmw_3_n_5}. The identified constitutive behavior exhibits a clear physical structure. The equilibrium branch shows significant activity in both \(\Psieqb_1(\Ioneb)\) and \(\Psieqb_2(\Itwob)\), with the latter being nearly linear and the former exhibiting more pronounced convexity. The corresponding final invariant windows are approximately
\[
\mathcal{I}(\Psieqb_1) \approx [3,10],
\qquad
\mathcal{I}(\Psieqb_2) \approx [0,10].
\]

These values are consistent with the maximum invariant values induced by the loading. For incompressible uniaxial tension,
\begin{equation}
    \Ioneb = \lambda^2 + 2\lambda^{-1},
    \qquad
    \Itwob = \left[2\lambda + \lambda^{-2}\right]^{3/2} - 3\sqrt{3}.
\end{equation}

Insertion of \(\lambda_{\max}=3\) yields
\[
\Ioneb \approx 9.6667,
\qquad
\Itwob \approx 9.9109,
\]
which agrees very well with the identified equilibrium windows.

The non-equilibrium branches display smaller energy magnitudes and reduced invariant windows. In particular, the second Maxwell branch operates over
\[
\mathcal{I}(\Psineqb_{1,2}) \approx [3,6],
\qquad
\mathcal{I}(\Psineqb_{2,2}) \approx [0,5],
\]
whereas for the first Maxwell branch the corresponding windows are much narrower,
\[
\mathcal{I}(\Psineqb_{1,1}) \approx [3,3.02],
\qquad
\mathcal{I}(\Psineqb_{2,1}) \approx [0,0.05].
\]

This indicates that the first Maxwell branch captures highly localized viscous effects, while the second Maxwell branch contributes over a broader, yet still distinctly non-equilibrium, range of invariants.

This behavior is consistent with the identified dissipation potentials. A steeper slope of \(\Phic_{\mw}(\Jtau{\mw})\) implies stronger viscous relaxation and therefore larger deviations from the elastic trial state in the local branch update. Conversely, a flatter dissipation potential corresponds to more elastic behavior and broader invariant windows. In the present study, the dissipation potential of the Maxwell branch \(2\) is steeper than that of branch \(1\), consistent with the observed separation of roles.

The evolution of the spline domains over the outer iterations is shown in~\cref{fig:Nmw_3_n_5}d. The dashed lines represent the log-sum-exp targets, while the solid lines denote the actual domain endpoints. The final windows reported above correspond to the state after the seven outer iterations and the final refinement phase. The results show that the domain endpoints align rapidly with the activated invariant range while remaining stable under the outer updates.

\begin{figure}
    \centering
    \includegraphics[width=1.0\linewidth]{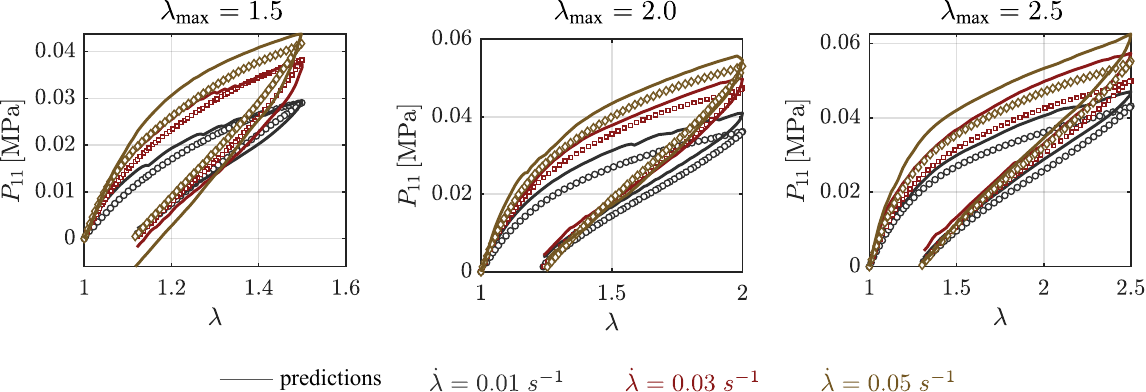}
    \caption{\textbf{Validation against unseen experimental data.} Comparison of the numerical predictions obtained with the learned constitutive functions from~\cref{fig:Nmw_3_n_5}b and the experimental loading--unloading data of VHB Tape 4910 from \cite{hossain_experimental_2012} (not included during calibration) for different maximum stretches $\lambda_{\text{max}} = \{1.5, 2.0, 2.5\}$ and stretch rates $\dot{\lambda}_{\text{max}} = \{0.01, 0.03, 0.05\} \; \rm{s}^{-1}.$}
    \label{fig:Nmw_3_n_5_pred}
\end{figure}

The model predictions show excellent agreement with the experimental data across both loading rates included in the calibration, with mean squared errors of \(\num{4.28e-7}\) for \(\dot{\lambda}=\SI{0.05}{\per\second}\) and \(\num{1.74e-7}\) for \(\dot{\lambda}=\SI{0.01}{\per\second}\). The model predictions for training data not included in the calibration set are shown in~\cref{fig:Nmw_3_n_5_pred}, indicating a fairly good match to the experimental data. Deviations are observed for $\lambda_{\mathrm{max}} = 2.5$, which, however, has already been reported by other authors, e.g., \cite{kalina_physics-augmented_2026}.

\begin{figure}
    \centering
    \includegraphics[width=0.85\linewidth]{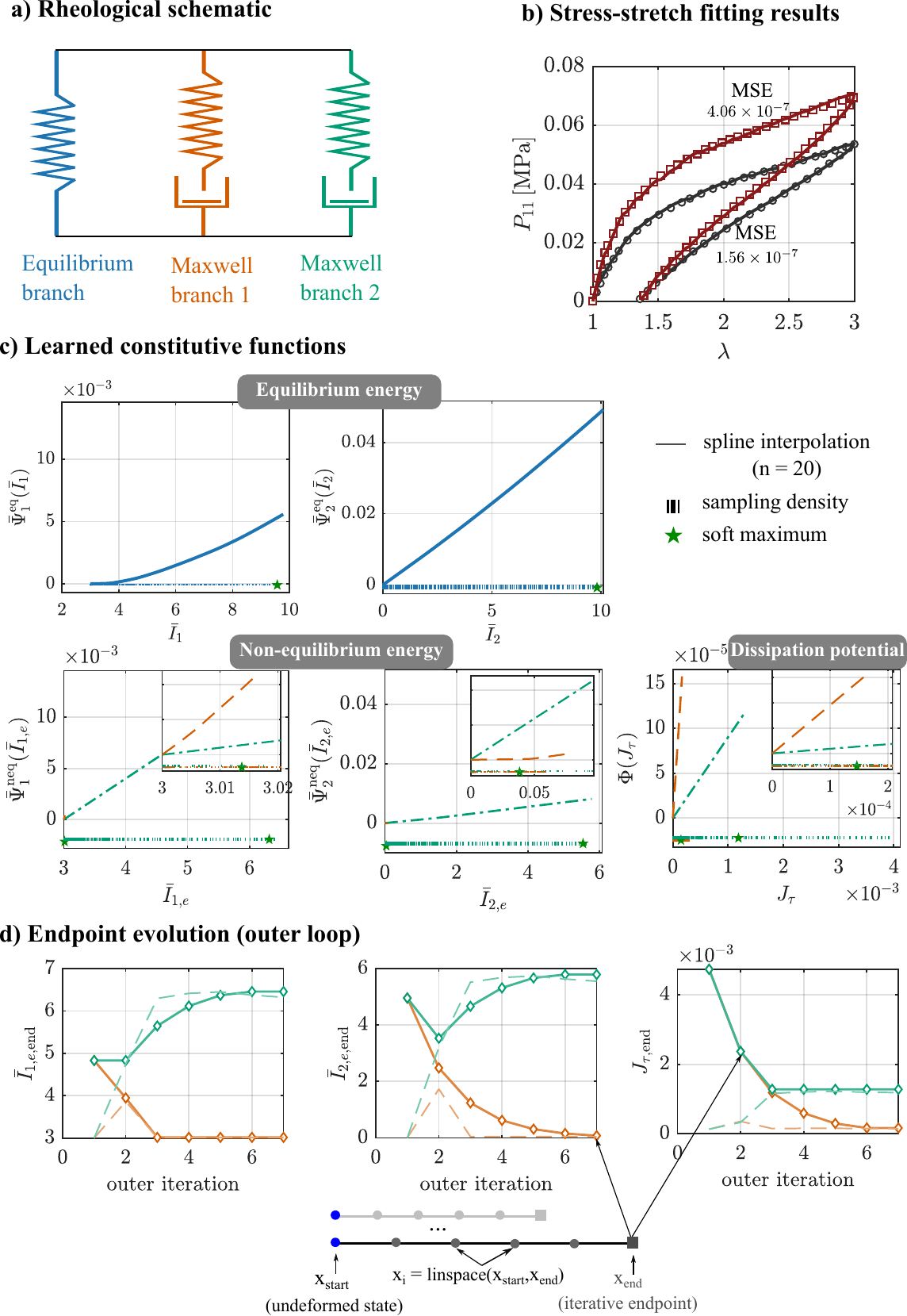}
    \caption{\textbf{Calibration results for the same rheological structure as in \cref{fig:Nmw_3_n_5}, but with \(n=20\) spline coefficients per scalar constitutive function.} (a) Rheological schematic. (b) Stress--stretch curves for both datasets together with the identified model response. (c) Identified constitutive spline functions and sampled invariant values. (d) Evolution of the domain endpoints over the outer iterations. Despite the increased spline resolution, the identified constitutive behavior remains essentially unchanged. Colors introduced in panel (a) consistently identify the branches of the rheological model and their associated contributions throughout the figure.}
    \label{fig:Nmw_3_n_20}
\end{figure}

Importantly, the results are convincingly insensitive to the number of spline coefficients. Increasing the spline resolution from \(n=5\) to \(n=20\) yields nearly identical constitutive responses, as shown in~\cref{fig:Nmw_3_n_20}. Although the reader may notice that the two non-equilibrium branches appear interchanged, this is not problematic: the generalized Maxwell response depends only on the sum of all branch contributions. Different discretizations may therefore lead to a different allocation of roles among the branches without changing the overall constitutive behavior.

The stability of both the constitutive functions and their invariant windows indicates that the proposed formulation yields a robust and well-conditioned identification problem.

\subsection{Enhanced DAVIS scales to highly overparameterized models}

In the previous subsection, we observed that a comparatively small model with two Maxwell elements and \(n=5\) spline coefficients per function already provides an excellent fit to the data. We now increase the model complexity to assess the scalability and robustness of the proposed formulation under strong overparameterization.

\begin{figure}
    \centering
    \includegraphics[width=0.85\linewidth]{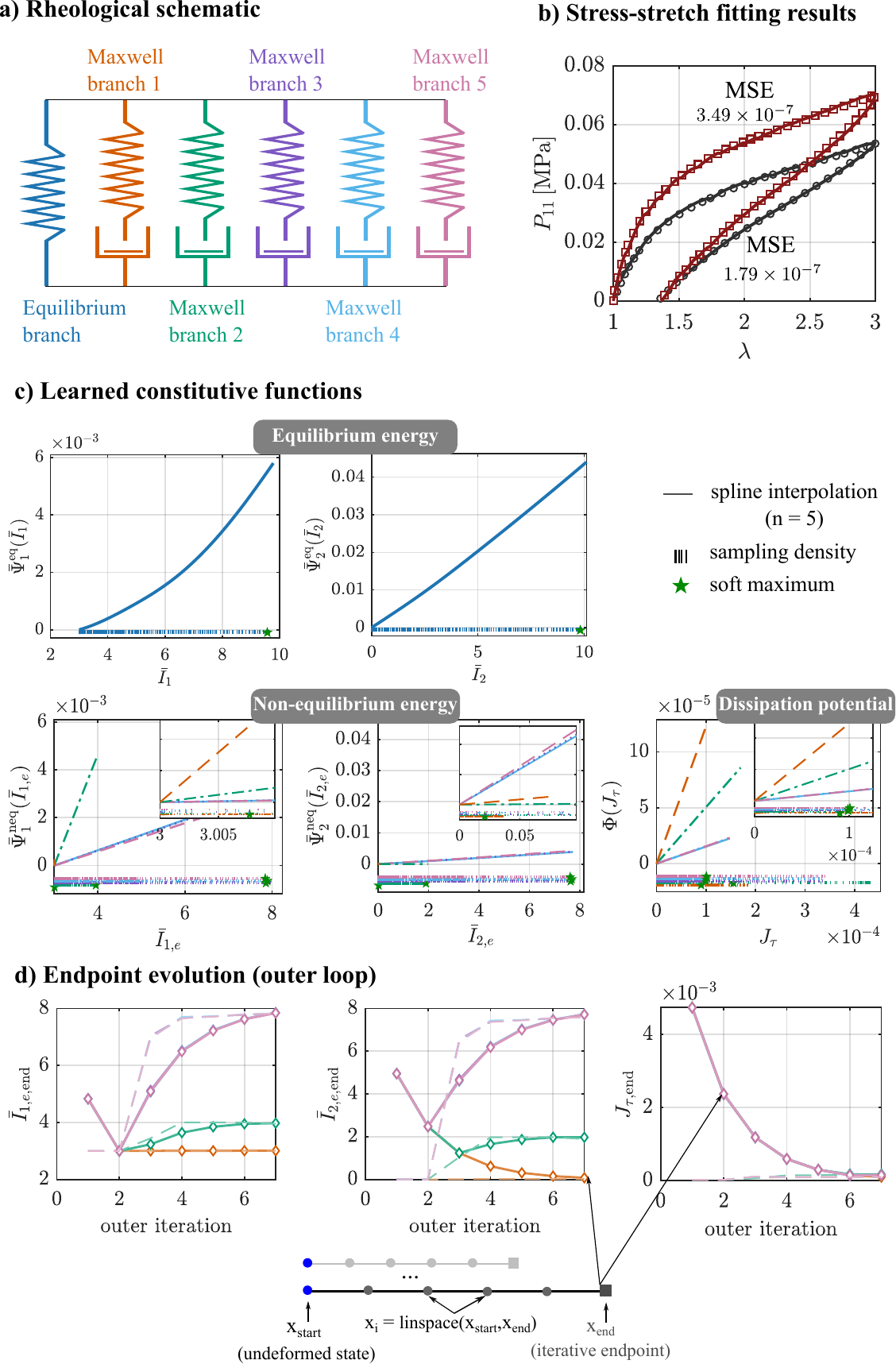}
    \caption{\textbf{Calibration results for a model with five Maxwell elements and \(n=5\) spline coefficients per scalar constitutive function.} (a) Rheological schematic. (b) Stress--stretch curves for both datasets together with the identified model response. (c) Identified constitutive spline functions; rug plots indicate the sampled invariant values and markers denote the soft-maximum targets for the adaptive domain endpoints. (d) Evolution of the domain endpoints \(x_{\mathrm{end}}^{(f)}\) over the outer iterations. Colors introduced in panel (a) consistently identify the branches of the rheological model and their associated contributions throughout the figure.}
    \label{fig:Nmw_6_n_5}
\end{figure}

First, we consider a model with five Maxwell elements and \(n=5\) spline coefficients per scalar constitutive function. Compared to the two-branch model, the constitutive response is distributed across multiple Maxwell elements, indicating redundancy in the representation. The results are presented in~\cref{fig:Nmw_6_n_5}.

The equilibrium branch remains essentially unchanged compared to the simpler model. In particular, the two equilibrium functions \(\Psieqb_1(\Ioneb)\) and \(\Psieqb_2(\Itwob)\) are identified very similarly to the two-branch case, with \(\Psieqb_2\) remaining nearly linear and \(\Psieqb_1\) retaining its more pronounced convexity. This indicates that the equilibrium response is stably identified by the data and is largely unaffected by the additional non-equilibrium branches.

The interpretation of the non-equilibrium branches becomes less straightforward in the overparameterized setting, since the response is now distributed across five Maxwell elements. Nevertheless, a consistent pattern remains visible. One branch again operates over a very narrow invariant range and therefore captures the strongly viscous effects. The remaining branches exhibit broader windows and more elastic-like behavior. In all cases, the invariant windows of the non-equilibrium energies remain smaller than those of the equilibrium branch, which is consistent with the role of the dissipation potentials: unless a branch behaves purely elastically, viscous evolution reduces the range of elastic invariants sampled by that branch. This point is particularly transparent in the present spline-based setting, since the invariant windows are identified explicitly as part of the constitutive representation. In contrast, such information is often not directly accessible in black-box neural network models.

\begin{figure}
    \centering
    \includegraphics[width=0.85\linewidth]{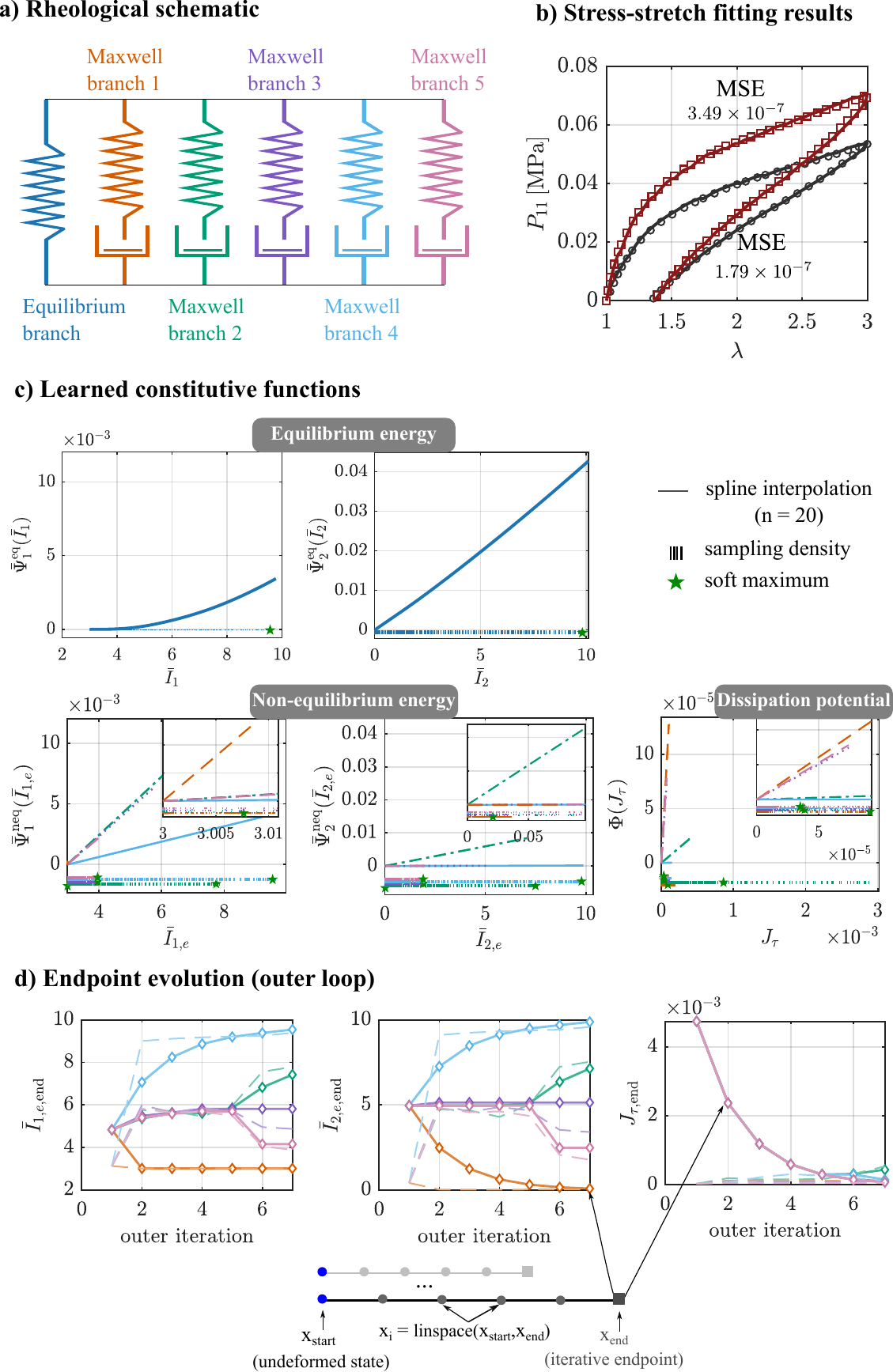}
    \caption{\textbf{Calibration results for a highly overparameterized model with five Maxwell elements and \(n=20\) spline coefficients per scalar constitutive function.} (a) Rheological schematic. (b) Stress--stretch curves for both datasets together with the identified response. (c) Identified constitutive spline functions and sampled invariant values. (d) Evolution of the adaptive domain endpoints. The macroscopic response remains stable despite the high-dimensional parameter space. Colors introduced in panel (a) consistently identify the branches of the rheological model and their associated contributions throughout the figure.}
    \label{fig:Nmw_6_n_20}
\end{figure}

Next, we increase the spline resolution to \(n=20\) while maintaining five Maxwell elements, resulting in a model with more than 300 optimization parameters. The results are shown in~\cref{fig:Nmw_6_n_20}.

The identified constitutive behavior remains consistent with the lower-dimensional cases. Again, the equilibrium branch is identified very similarly to the \(n=5\) model, and the non-equilibrium spline functions remain of comparable magnitude. One branch carries the strongly viscous effects through very narrow windows in \(\Ionee{\mw}\) and \(\Itwoe{\mw}\), whereas the remaining branches are more elastic-like and operate on broader windows. This separation is again reflected in the identified dissipation potentials, which exhibit different slopes from flatter to steeper profiles.

While the interpretation of individual branch contributions becomes less unique in this regime, the overall response remains stable and physically meaningful. These findings highlight the scalability of the proposed framework to high-dimensional identification problems.

\subsection{Group-wise sparsity identifies a parsimonious Maxwell structure}

In the previous subsections, the number of Maxwell elements was fixed and the calibration was performed without sparsity promotion. The purpose of those studies was to demonstrate that the proposed formulation remains robust for different spline resolutions and overall model complexities. We now turn to the sparsity-promoting extension in order to address model selection.

Starting from an overparameterized model with five Maxwell elements, we activate the sparsity term and vary \(\lambda_{\mathrm{sparse}}\) over several orders of magnitude. Each run is initialized from the previous solution in order to ensure continuity along the sparsity path and to improve numerical efficiency. For each value of \(\lambda_{\mathrm{sparse}}\), the data misfit is evaluated separately from the sparsity term. In addition, branch activity is assessed using the measure
\[
A_{\mw}
=
\left[
A_{1,\mw}^{2}
+
A_{2,\mw}^{2}
+
A_{\Phi,\mw}^{2}
\right]^{1/2},
\]
and a branch is classified as active if
\[
\frac{A_{\mw}}{\max_{k} A_{k}} > \varepsilon_{\mathrm{act}},
\qquad
\varepsilon_{\mathrm{act}} = 10^{-3}.
\]

\begin{figure}
    \centering
    \includegraphics[width=0.98\linewidth]{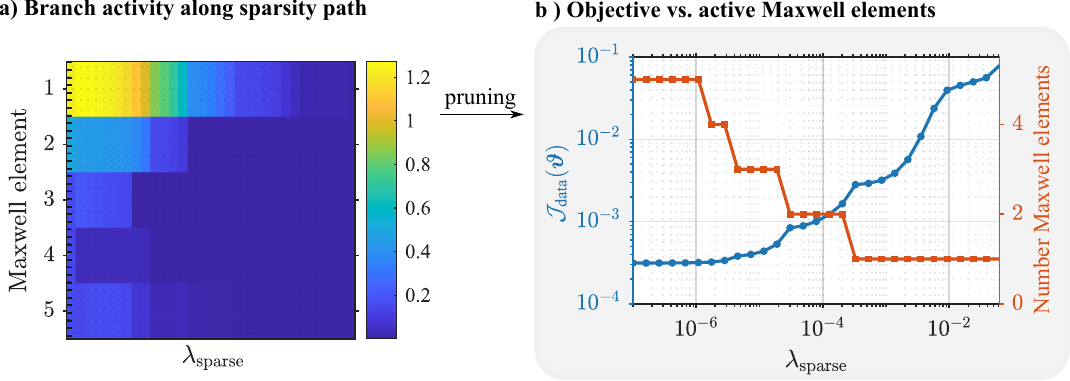}
    \caption{\textbf{Sparsity analysis.} (a) Heat map of the branch activity measures along the sparsity path for all five Maxwell elements. (b) Data loss \(\mathcal{J}_{\mathrm{data}}\) as a function of \(\lambda_{\mathrm{sparse}}\) together with the corresponding number of active Maxwell branches.}
    \label{fig:fig_sparsity}
\end{figure}

The results are shown in~\cref{fig:fig_sparsity}. For small values of \(\lambda_{\mathrm{sparse}}\), all branches remain active, indicating that the regularization is too weak to suppress redundant contributions. As \(\lambda_{\mathrm{sparse}}\) increases, the activity becomes concentrated in fewer branches, and the less relevant Maxwell elements are progressively pruned.

The right panel of~\cref{fig:fig_sparsity} shows the corresponding trade-off between the misfit of the data and the complexity of the model. For very small values of \(\lambda_{\mathrm{sparse}}\), the data loss remains close to its minimum while several Maxwell branches are active. Increasing \(\lambda_{\mathrm{sparse}}\) first reduces the number of active branches with only a moderate change in data mismatch. However, beyond this regime, further pruning leads to a rapid deterioration of the fit.

Based on this trade-off, we select $\lambda_{\mathrm{sparse}} = 10^{-4}$, for which the model reduces to two active non-equilibrium Maxwell branches while still retaining a data misfit close to the minimum achieved by the fully populated model. This confirms that two Maxwell elements provide a favorable compromise between accuracy and parsimony for the present dataset.

\section{Conclusion}
\label{sec:conclusion}

In this work, we revisited the data-adaptive spline-based (DAVIS) framework for finite strain viscoelasticity and identified key sources of non-robustness in its original formulation. In particular, the combination of value-based spline representation and the simultaneous optimization of spline values and domain endpoints was shown to introduce structural ambiguities and ill-conditioning.
To address these issues, two modifications were proposed. First, the spline representation was reformulated in terms of a curvature-based parameterization, in which convexity and monotonicity are enforced by construction through non-negative curvature coefficients and boundary slopes. Second, the adaptation of the interpolation domains was decoupled from the inner parameter identification and treated within a block-alternating scheme driven by invariant samples obtained from the forward solution.

Enhanced DAVIS is a well-conditioned and robust identification problem. Our method was shown to be largely insensitive to the spline resolution and to remain stable even in highly overparameterized settings involving multiple Maxwell branches and several hundred parameters. Multi-start studies demonstrated convergence to a unique solution across a wide range of initializations, in contrast to the original DAVIS approach. 

Beyond robustness, the spline-based representation provides direct insight into the constitutive structure through the identified invariant windows. These windows reveal a clear separation of roles among the Maxwell branches, with strongly viscous effects confined to narrow invariant ranges and more elastic contributions acting over broader domains. This level of interpretability is a distinctive advantage of the proposed approach.

Finally, the robustness of the formulation enables the incorporation of a group-wise sparsity regularization, which allows for the automatic identification of a parsimonious Maxwell structure. The resulting models achieve a favorable balance between accuracy and complexity without requiring a priori selection of the number of branches.

\section*{Code availability}
The code for the calibration is publicly available in \url{https://github.com/swiesheier/davis-viscoelasticity-robust}.

\section*{Acknowledgments}
Simon Wiesheier, Paul Steinmann, and Miguel Angel Moreno-Mateos acknowledge support from the European Research Council (ERC) under the Horizon Europe program (Grant-No. 101052785, project: SoftFrac). Funded by the European Union. Views and opinions expressed are, however, those of the author(s) only and do not necessarily reflect those of the European Union or the European Research Council Executive Agency. Neither the European Union nor the granting authority can be held responsible for them.

\begin{figure}[ht]
\includegraphics[width=0.3\textwidth]{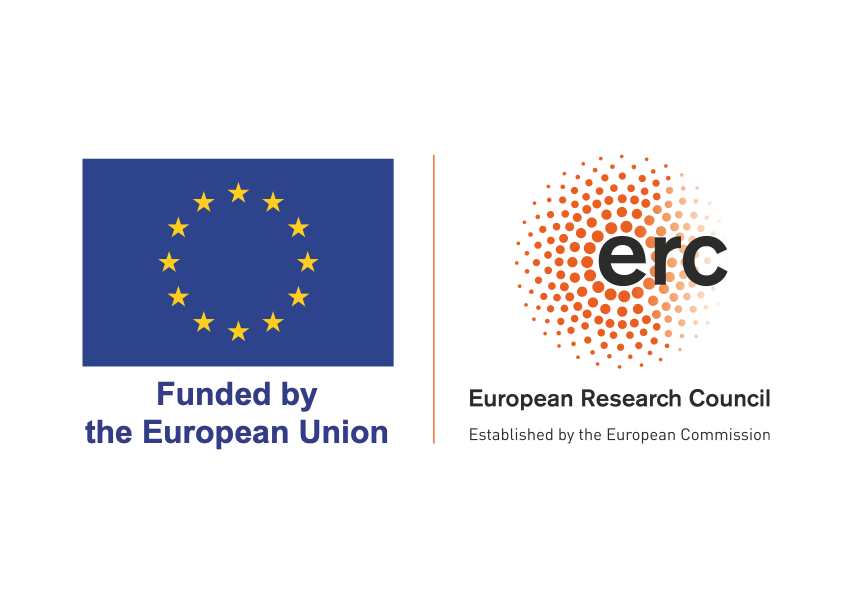}
\end{figure}

\section*{Competing Interests}
\noindent The Authors declare no Competing Financial or Non-Financial Interests.


\bibliographystyle{naturemag}

\newpage

\addcontentsline{toc}{section}{References}

\begin{thebibliography}{10}
\expandafter\ifx\csname url\endcsname\relax
  \def\url#1{\texttt{#1}}\fi
\expandafter\ifx\csname urlprefix\endcsname\relax\def\urlprefix{URL }\fi
\providecommand{\bibinfo}[2]{#2}
\providecommand{\eprint}[2][]{\url{#2}}

\bibitem{steinmann_hyperelastic_2012}
\bibinfo{author}{Steinmann, P.}, \bibinfo{author}{Hossain, M.} \&
  \bibinfo{author}{Possart, G.}
\newblock \bibinfo{title}{Hyperelastic models for rubber-like materials:
  consistent tangent operators and suitability for {Treloar}’s data}.
\newblock \emph{\bibinfo{journal}{Archive of Applied Mechanics}}
  \textbf{\bibinfo{volume}{82}}, \bibinfo{pages}{1183--1217}
  (\bibinfo{year}{2012}).

\bibitem{ricker_systematic_2023}
\bibinfo{author}{Ricker, A.} \& \bibinfo{author}{Wriggers, P.}
\newblock \bibinfo{title}{Systematic {Fitting} and {Comparison} of
  {Hyperelastic} {Continuum} {Models} for {Elastomers}}.
\newblock \emph{\bibinfo{journal}{Archives of Computational Methods in
  Engineering}} \textbf{\bibinfo{volume}{30}} (\bibinfo{year}{2023}).

\bibitem{holzapfel_nonlinear_2001}
\bibinfo{author}{Holzapfel, G.}
\newblock \emph{\bibinfo{title}{Nonlinear solid mechanics. {A} {Continuum}
  {Approach} for {Engineering}}} (\bibinfo{publisher}{John Wiley \& Sons},
  \bibinfo{year}{2001}), \bibinfo{edition}{second print} edn.

\bibitem{le_tallec_three-dimensional_1993}
\bibinfo{author}{Le~Tallec, P.}, \bibinfo{author}{Rahier, C.} \&
  \bibinfo{author}{Kaiss, A.}
\newblock \bibinfo{title}{Three-dimensional incompressible viscoelasticity in
  large strains: {Formulation} and numerical approximation}.
\newblock \emph{\bibinfo{journal}{Computer Methods in Applied Mechanics and
  Engineering}} \textbf{\bibinfo{volume}{109}}, \bibinfo{pages}{233--258}
  (\bibinfo{year}{1993}).

\bibitem{lion_physically_1997}
\bibinfo{author}{Lion, A.}
\newblock \bibinfo{title}{A physically based method to represent the
  thermo-mechanical behaviour of elastomers}.
\newblock \emph{\bibinfo{journal}{Acta Mechanica}}
  \textbf{\bibinfo{volume}{123}}, \bibinfo{pages}{1--25}
  (\bibinfo{year}{1997}).

\bibitem{reese_theory_1998}
\bibinfo{author}{Reese, S.} \& \bibinfo{author}{Govindjee, S.}
\newblock \bibinfo{title}{A theory of finite viscoelasticity and numerical
  aspects}.
\newblock \emph{\bibinfo{journal}{International Journal of Solids and
  Structures}} \textbf{\bibinfo{volume}{35}}, \bibinfo{pages}{3455--3482}
  (\bibinfo{year}{1998}).

\bibitem{lubliner_model_1985}
\bibinfo{author}{Lubliner, J.}
\newblock \bibinfo{title}{A model of rubber viscoelasticity}.
\newblock \emph{\bibinfo{journal}{Mechanics Research Communications}}
  \textbf{\bibinfo{volume}{12}}, \bibinfo{pages}{93--99}
  (\bibinfo{year}{1985}).

\bibitem{govindjee_presentation_1997}
\bibinfo{author}{Govindjee, S.} \& \bibinfo{author}{Reese, S.}
\newblock \bibinfo{title}{A {Presentation} and {Comparison} of {Two} {Large}
  {Deformation} {Viscoelasticity} {Models}}.
\newblock \emph{\bibinfo{journal}{Journal of Engineering Materials and
  Technology}} \textbf{\bibinfo{volume}{119}}, \bibinfo{pages}{251--255}
  (\bibinfo{year}{1997}).

\bibitem{gouhier_comparison_2024}
\bibinfo{author}{Gouhier, F.} \& \bibinfo{author}{Diani, J.}
\newblock \bibinfo{title}{A comparison of finite strain viscoelastic models
  based on the multiplicative decomposition}.
\newblock \emph{\bibinfo{journal}{European Journal of Mechanics - A/Solids}}
  \textbf{\bibinfo{volume}{108}}, \bibinfo{pages}{105424}
  (\bibinfo{year}{2024}).

\bibitem{kirchdoerfer_data-driven_2016}
\bibinfo{author}{Kirchdoerfer, T.} \& \bibinfo{author}{Ortiz, M.}
\newblock \bibinfo{title}{Data-driven computational mechanics}.
\newblock \emph{\bibinfo{journal}{Computer Methods in Applied Mechanics and
  Engineering}} \textbf{\bibinfo{volume}{304}}, \bibinfo{pages}{81--101}
  (\bibinfo{year}{2016}).

\bibitem{fuhg_review_2025}
\bibinfo{author}{Fuhg, J.~N.} \emph{et~al.}
\newblock \bibinfo{title}{A {Review} on {Data}-{Driven} {Constitutive} {Laws}
  for {Solids}}.
\newblock \emph{\bibinfo{journal}{Archives of Computational Methods in
  Engineering}} \textbf{\bibinfo{volume}{32}}, \bibinfo{pages}{1841--1883}
  (\bibinfo{year}{2025}).

\bibitem{kalina_automated_2022}
\bibinfo{author}{Kalina, K.~A.}, \bibinfo{author}{Linden, L.},
  \bibinfo{author}{Brummund, J.}, \bibinfo{author}{Metsch, P.} \&
  \bibinfo{author}{Kästner, M.}
\newblock \bibinfo{title}{Automated constitutive modeling of isotropic
  hyperelasticity based on artificial neural networks}.
\newblock \emph{\bibinfo{journal}{Computational Mechanics}}
  \textbf{\bibinfo{volume}{69}}, \bibinfo{pages}{1--20} (\bibinfo{year}{2022}).

\bibitem{klein_polyconvex_2022}
\bibinfo{author}{Klein, D.~K.}, \bibinfo{author}{Fernández, M.},
  \bibinfo{author}{Martin, R.~J.}, \bibinfo{author}{Neff, P.} \&
  \bibinfo{author}{Weeger, O.}
\newblock \bibinfo{title}{Polyconvex anisotropic hyperelasticity with neural
  networks}.
\newblock \emph{\bibinfo{journal}{Journal of the Mechanics and Physics of
  Solids}} \textbf{\bibinfo{volume}{159}}, \bibinfo{pages}{104703}
  (\bibinfo{year}{2022}).

\bibitem{klein_neural_2025}
\bibinfo{author}{Klein, D.~K.} \emph{et~al.}
\newblock \bibinfo{title}{Neural networks meet hyperelasticity: {A} monotonic
  approach} (\bibinfo{year}{2025}).
\newblock \bibinfo{note}{ArXiv:2501.02670 [cs]}.

\bibitem{linden_neural_2023}
\bibinfo{author}{Linden, L.} \emph{et~al.}
\newblock \bibinfo{title}{Neural networks meet hyperelasticity: {A} guide to
  enforcing physics}.
\newblock \emph{\bibinfo{journal}{Journal of the Mechanics and Physics of
  Solids}} \textbf{\bibinfo{volume}{179}}, \bibinfo{pages}{105363}
  (\bibinfo{year}{2023}).

\bibitem{thakolkaran_can_2025}
\bibinfo{author}{Thakolkaran, P.} \emph{et~al.}
\newblock \bibinfo{title}{Can {KAN} {CANs}? {Input}-convex
  {Kolmogorov}-{Arnold} {Networks} ({KANs}) as hyperelastic constitutive
  artificial neural networks ({CANs})}.
\newblock \emph{\bibinfo{journal}{Computer Methods in Applied Mechanics and
  Engineering}} \textbf{\bibinfo{volume}{443}}, \bibinfo{pages}{118089}
  (\bibinfo{year}{2025}).

\bibitem{abdolazizi_constitutive_2025}
\bibinfo{author}{Abdolazizi, K.~P.}, \bibinfo{author}{Aydin, R.~C.},
  \bibinfo{author}{Cyron, C.~J.} \& \bibinfo{author}{Linka, K.}
\newblock \bibinfo{title}{Constitutive {Kolmogorov}–{Arnold} {Networks}
  ({CKANs}): {Combining} accuracy and interpretability in data-driven material
  modeling}.
\newblock \emph{\bibinfo{journal}{Journal of the Mechanics and Physics of
  Solids}} \textbf{\bibinfo{volume}{203}}, \bibinfo{pages}{106212}
  (\bibinfo{year}{2025}).

\bibitem{frankel_tensor_2019}
\bibinfo{author}{Frankel, A.~L.}, \bibinfo{author}{Jones, R.~E.} \&
  \bibinfo{author}{Swiler, L.~P.}
\newblock \bibinfo{title}{Tensor {Basis} {Gaussian} {Process} {Models} of
  {Hyperelastic} {Materials}}.
\newblock \emph{\bibinfo{journal}{ArXiv}}  (\bibinfo{year}{2019}).
\newblock \bibinfo{note}{Abs/1912.10872}.

\bibitem{abdusalamov_automatic_2023}
\bibinfo{author}{Abdusalamov, R.}, \bibinfo{author}{Hillgärtner, M.} \&
  \bibinfo{author}{Itskov, M.}
\newblock \bibinfo{title}{Automatic generation of interpretable hyperelastic
  material models by symbolic regression}.
\newblock \emph{\bibinfo{journal}{International Journal for Numerical Methods
  in Engineering}} \textbf{\bibinfo{volume}{124}}, \bibinfo{pages}{2093--2104}
  (\bibinfo{year}{2023}).

\bibitem{tac_data-driven_2023}
\bibinfo{author}{Taç, V.}, \bibinfo{author}{Rausch, M.~K.},
  \bibinfo{author}{Sahli~Costabal, F.} \& \bibinfo{author}{Tepole, A.~B.}
\newblock \bibinfo{title}{Data-driven anisotropic finite viscoelasticity using
  neural ordinary differential equations}.
\newblock \emph{\bibinfo{journal}{Computer Methods in Applied Mechanics and
  Engineering}} \textbf{\bibinfo{volume}{411}}, \bibinfo{pages}{116046}
  (\bibinfo{year}{2023}).

\bibitem{flaschel_automated_2023}
\bibinfo{author}{Flaschel, M.}, \bibinfo{author}{Kumar, S.} \&
  \bibinfo{author}{De~Lorenzis, L.}
\newblock \bibinfo{title}{Automated discovery of generalized standard material
  models with {EUCLID}}.
\newblock \emph{\bibinfo{journal}{Computer Methods in Applied Mechanics and
  Engineering}} \textbf{\bibinfo{volume}{405}}, \bibinfo{pages}{115867}
  (\bibinfo{year}{2023}).

\bibitem{flaschel_unsupervised_2026}
\bibinfo{author}{Flaschel, M.}, \bibinfo{author}{Moreno-Mateos, M.~A.},
  \bibinfo{author}{Wiesheier, S.}, \bibinfo{author}{Steinmann, P.} \&
  \bibinfo{author}{Kuhl, E.}
\newblock \bibinfo{title}{Unsupervised {Material} {Fingerprinting}:
  {Ultra}-fast hyperelastic model discovery from full-field experimental
  measurements} (\bibinfo{year}{2026}).
\newblock \bibinfo{note}{Version Number: 1}.

\bibitem{sussman_model_2009}
\bibinfo{author}{Sussman, T.} \& \bibinfo{author}{Bathe, K.-J.}
\newblock \bibinfo{title}{A model of incompressible isotropic hyperelastic
  material behavior using spline interpolations of tension–compression test
  data}.
\newblock \emph{\bibinfo{journal}{Communications in numerical methods in
  engineering}} \textbf{\bibinfo{volume}{25}}, \bibinfo{pages}{53--63}
  (\bibinfo{year}{2009}).

\bibitem{dal_data-driven_2023}
\bibinfo{author}{Dal, H.}, \bibinfo{author}{Denli, F.~A.},
  \bibinfo{author}{Açan, A.~K.} \& \bibinfo{author}{Kaliske, M.}
\newblock \bibinfo{title}{Data-driven hyperelasticity, {Part} {I}: {A}
  canonical isotropic formulation for rubberlike materials}.
\newblock \emph{\bibinfo{journal}{Journal of the Mechanics and Physics of
  Solids}} \textbf{\bibinfo{volume}{179}}, \bibinfo{pages}{105381}
  (\bibinfo{year}{2023}).

\bibitem{tikenogullari_data-driven_2023}
\bibinfo{author}{Tikenoğulları, O.~Z.}, \bibinfo{author}{Açan, A.~K.},
  \bibinfo{author}{Kuhl, E.} \& \bibinfo{author}{Dal, H.}
\newblock \bibinfo{title}{Data-driven hyperelasticity, {Part} {II}: {A}
  canonical framework for anisotropic soft biological tissues}.
\newblock \emph{\bibinfo{journal}{Journal of the Mechanics and Physics of
  Solids}} \textbf{\bibinfo{volume}{181}}, \bibinfo{pages}{105453}
  (\bibinfo{year}{2023}).

\bibitem{wiesheier_versatile_2024}
\bibinfo{author}{Wiesheier, S.}, \bibinfo{author}{Moreno-Mateos, M.~A.} \&
  \bibinfo{author}{Steinmann, P.}
\newblock \bibinfo{title}{Versatile data-adaptive hyperelastic energy functions
  for soft materials}.
\newblock \emph{\bibinfo{journal}{Computer Methods in Applied Mechanics and
  Engineering}} \textbf{\bibinfo{volume}{430}}, \bibinfo{pages}{117208}
  (\bibinfo{year}{2024}).

\bibitem{moreno-mateos_biaxial_2025}
\bibinfo{author}{Moreno-Mateos, M.~A.}, \bibinfo{author}{Wiesheier, S.},
  \bibinfo{author}{Esmaeili, A.}, \bibinfo{author}{Hossain, M.} \&
  \bibinfo{author}{Steinmann, P.}
\newblock \bibinfo{title}{Biaxial characterization of soft elastomers:
  experiments and data-adaptive configurational forces for fracture}
  (\bibinfo{year}{2025}).
\newblock \bibinfo{note}{ArXiv:2505.20244 [cond-mat]}.

\bibitem{wiesheier_data-adaptive_2026}
\bibinfo{author}{Wiesheier, S.}, \bibinfo{author}{Moreno-Mateos, M.~A.} \&
  \bibinfo{author}{Steinmann, P.}
\newblock \bibinfo{title}{Data-adaptive spline surfaces for non-separable
  hyperelastic energy functions} (\bibinfo{year}{2026}).
\newblock \bibinfo{note}{ArXiv:2604.10059 [cs]}.

\bibitem{halphen_sur_1975}
\bibinfo{author}{Halphen, B.} \& \bibinfo{author}{Nguyen, Q.~S.}
\newblock \bibinfo{title}{Sur les matériaux standard généralisés}.
\newblock \emph{\bibinfo{journal}{Journal de Mécanique}}
  \textbf{\bibinfo{volume}{14}}, \bibinfo{pages}{39--63}
  (\bibinfo{year}{1975}).

\bibitem{ziegler_thermodynamik_1957}
\bibinfo{author}{Ziegler, H.}
\newblock \bibinfo{title}{Thermodynamik und rheologische {Probleme}}.
\newblock \emph{\bibinfo{journal}{Ingenieur-Archiv}}
  \textbf{\bibinfo{volume}{25}}, \bibinfo{pages}{58--70}
  (\bibinfo{year}{1957}).

\bibitem{ziegler_attempt_1958}
\bibinfo{author}{Ziegler, H.}
\newblock \bibinfo{title}{An attempt to generalize {Onsager}'s principle, and
  its significance for rheological problems}.
\newblock \emph{\bibinfo{journal}{ZAMP Zeitschrift für Angewandte Mathematik
  und Physik}} \textbf{\bibinfo{volume}{9}}, \bibinfo{pages}{748--763}
  (\bibinfo{year}{1958}).

\bibitem{kumar_two-potential_2016}
\bibinfo{author}{Kumar, A.} \& \bibinfo{author}{Lopez-Pamies, O.}
\newblock \bibinfo{title}{On the two-potential constitutive modeling of rubber
  viscoelastic materials}.
\newblock \emph{\bibinfo{journal}{Comptes Rendus. Mécanique}}
  \textbf{\bibinfo{volume}{344}}, \bibinfo{pages}{102--112}
  (\bibinfo{year}{2016}).

\bibitem{yu_onsagernet_2021}
\bibinfo{author}{Yu, H.}, \bibinfo{author}{Tian, X.}, \bibinfo{author}{E, W.}
  \& \bibinfo{author}{Li, Q.}
\newblock \bibinfo{title}{{OnsagerNet}: {Learning} stable and interpretable
  dynamics using a generalized {Onsager} principle}.
\newblock \emph{\bibinfo{journal}{Physical Review Fluids}}
  \textbf{\bibinfo{volume}{6}}, \bibinfo{pages}{114402} (\bibinfo{year}{2021}).

\bibitem{huang_variational_2022}
\bibinfo{author}{Huang, S.}, \bibinfo{author}{He, Z.}, \bibinfo{author}{Chem,
  B.} \& \bibinfo{author}{Reina, C.}
\newblock \bibinfo{title}{Variational {Onsager} {Neural} {Networks} ({VONNs}):
  {A} thermodynamics-based variational learning strategy for non-equilibrium
  {PDEs}}.
\newblock \emph{\bibinfo{journal}{Journal of the Mechanics and Physics of
  Solids}} \textbf{\bibinfo{volume}{163}}, \bibinfo{pages}{104856}
  (\bibinfo{year}{2022}).

\bibitem{flaschel_automated_2023-1}
\bibinfo{author}{Flaschel, M.}, \bibinfo{author}{Kumar, S.} \&
  \bibinfo{author}{De~Lorenzis, L.}
\newblock \bibinfo{title}{Automated discovery of generalized standard material
  models with {EUCLID}}.
\newblock \emph{\bibinfo{journal}{Computer Methods in Applied Mechanics and
  Engineering}} \textbf{\bibinfo{volume}{405}}, \bibinfo{pages}{115867}
  (\bibinfo{year}{2023}).

\bibitem{rosenkranz_viscoelasticty_2024}
\bibinfo{author}{Rosenkranz, M.}, \bibinfo{author}{Kalina, K.~A.},
  \bibinfo{author}{Brummund, J.}, \bibinfo{author}{Sun, W.} \&
  \bibinfo{author}{Kästner, M.}
\newblock \bibinfo{title}{Viscoelasticty with physics-augmented neural
  networks: model formulation and training methods without prescribed internal
  variables}.
\newblock \emph{\bibinfo{journal}{Computational Mechanics}}
  \textbf{\bibinfo{volume}{74}}, \bibinfo{pages}{1279--1301}
  (\bibinfo{year}{2024}).

\bibitem{kalina_physics-augmented_2026}
\bibinfo{author}{Kalina, K.~A.}, \bibinfo{author}{Brummund, J.} \&
  \bibinfo{author}{Kästner, M.}
\newblock \bibinfo{title}{A physics-augmented neural network framework for
  finite strain incompressible viscoelasticity}.
\newblock \emph{\bibinfo{journal}{Computer Methods in Applied Mechanics and
  Engineering}} \textbf{\bibinfo{volume}{455}}, \bibinfo{pages}{118892}
  (\bibinfo{year}{2026}).

\bibitem{holthusen_complement_2026}
\bibinfo{author}{Holthusen, H.} \& \bibinfo{author}{Kuhl, E.}
\newblock \bibinfo{title}{A complement to neural networks for anisotropic
  inelasticity at finite strains}.
\newblock \emph{\bibinfo{journal}{Computer Methods in Applied Mechanics and
  Engineering}} \textbf{\bibinfo{volume}{450}}, \bibinfo{pages}{118612}
  (\bibinfo{year}{2026}).

\bibitem{holthusen_generalized_2026}
\bibinfo{author}{Holthusen, H.}, \bibinfo{author}{Linka, K.},
  \bibinfo{author}{Kuhl, E.} \& \bibinfo{author}{Brepols, T.}
\newblock \bibinfo{title}{A generalized dual potential for inelastic
  {Constitutive} {Artificial} {Neural} {Networks}: {A} {JAX} implementation at
  finite strains}.
\newblock \emph{\bibinfo{journal}{Journal of the Mechanics and Physics of
  Solids}} \textbf{\bibinfo{volume}{206}}, \bibinfo{pages}{106337}
  (\bibinfo{year}{2026}).

\bibitem{upadhyay_physics-informed_2024}
\bibinfo{author}{Upadhyay, K.}, \bibinfo{author}{Fuhg, J.~N.},
  \bibinfo{author}{Bouklas, N.} \& \bibinfo{author}{Ramesh, K.~T.}
\newblock \bibinfo{title}{Physics-informed data-driven discovery of
  constitutive models with application to strain-rate-sensitive soft
  materials}.
\newblock \emph{\bibinfo{journal}{Computational Mechanics}}
  (\bibinfo{year}{2024}).

\bibitem{upadhyay_visco-hyperelastic_2020}
\bibinfo{author}{Upadhyay, K.}, \bibinfo{author}{Subhash, G.} \&
  \bibinfo{author}{Spearot, D.}
\newblock \bibinfo{title}{Visco-hyperelastic constitutive modeling of strain
  rate sensitive soft materials}.
\newblock \emph{\bibinfo{journal}{Journal of the Mechanics and Physics of
  Solids}} \textbf{\bibinfo{volume}{135}}, \bibinfo{pages}{103777}
  (\bibinfo{year}{2020}).

\bibitem{abdolazizi_viscoelastic_2024}
\bibinfo{author}{Abdolazizi, K.~P.}, \bibinfo{author}{Linka, K.} \&
  \bibinfo{author}{Cyron, C.~J.}
\newblock \bibinfo{title}{Viscoelastic constitutive artificial neural networks
  ({vCANNs}) – {A} framework for data-driven anisotropic nonlinear finite
  viscoelasticity}.
\newblock \emph{\bibinfo{journal}{Journal of Computational Physics}}
  \textbf{\bibinfo{volume}{499}}, \bibinfo{pages}{112704}
  (\bibinfo{year}{2024}).

\bibitem{wiesheier_data-adaptive_2026-1}
\bibinfo{author}{Wiesheier, S.}, \bibinfo{author}{Moreno-Mateos, M.~A.} \&
  \bibinfo{author}{Steinmann, P.}
\newblock \bibinfo{title}{Data-adaptive spline-based viscoelasticity for soft
  solids}.
\newblock \emph{\bibinfo{journal}{Computer Methods in Applied Mechanics and
  Engineering}} \textbf{\bibinfo{volume}{451}}, \bibinfo{pages}{118705}
  (\bibinfo{year}{2026}).

\bibitem{mahnken_unified_1996}
\bibinfo{author}{Mahnken, R.} \& \bibinfo{author}{Stein, E.}
\newblock \bibinfo{title}{A unified approach for parameter identification of
  inelastic material models in the frame of the finite element method}.
\newblock \emph{\bibinfo{journal}{Computer methods in applied mechanics and
  engineering}} \textbf{\bibinfo{volume}{136}}, \bibinfo{pages}{225--258}
  (\bibinfo{year}{1996}).

\bibitem{sutton_determination_1983}
\bibinfo{author}{Sutton, M.}, \bibinfo{author}{Wolters, W.},
  \bibinfo{author}{Peters, W.}, \bibinfo{author}{Ranson, W.} \&
  \bibinfo{author}{McNeill, S.}
\newblock \bibinfo{title}{Determination of displacements using an improved
  digital correlation method}.
\newblock \emph{\bibinfo{journal}{Image and vision computing}}
  \textbf{\bibinfo{volume}{1}}, \bibinfo{pages}{133--139}
  (\bibinfo{year}{1983}).

\bibitem{avril_overview_2008}
\bibinfo{author}{Avril, S.} \emph{et~al.}
\newblock \bibinfo{title}{Overview of {Identification} {Methods} of
  {Mechanical} {Parameters} {Based} on {Full}-field {Measurements}}.
\newblock \emph{\bibinfo{journal}{Experimental Mechanics}}
  \textbf{\bibinfo{volume}{48}}, \bibinfo{pages}{381--402}
  (\bibinfo{year}{2008}).

\bibitem{hartmann_identifiability_2018}
\bibinfo{author}{Hartmann, S.} \& \bibinfo{author}{Gilbert, R.}
\newblock \bibinfo{title}{Identifiability of material parameters in solid
  mechanics}.
\newblock \emph{\bibinfo{journal}{Archive of Applied Mechanics}}
  \textbf{\bibinfo{volume}{88}}, \bibinfo{pages}{3--26} (\bibinfo{year}{2018}).

\bibitem{hartmann_polyconvexity_2003}
\bibinfo{author}{Hartmann, S.} \& \bibinfo{author}{Neff, P.}
\newblock \bibinfo{title}{Polyconvexity of generalized polynomial-type
  hyperelastic strain energy functions for near-incompressibility}.
\newblock \emph{\bibinfo{journal}{International Journal of Solids and
  Structures}} \textbf{\bibinfo{volume}{40}}, \bibinfo{pages}{2767--2791}
  (\bibinfo{year}{2003}).

\bibitem{schroder_invariant_2003}
\bibinfo{author}{Schröder, J.} \& \bibinfo{author}{Neff, P.}
\newblock \bibinfo{title}{Invariant formulation of hyperelastic transverse
  isotropy based on polyconvex free energy functions}.
\newblock \emph{\bibinfo{journal}{International Journal of Solids and
  Structures}} \textbf{\bibinfo{volume}{40}}, \bibinfo{pages}{401--445}
  (\bibinfo{year}{2003}).

\bibitem{flory_thermodynamic_1961}
\bibinfo{author}{Flory, P.}
\newblock \bibinfo{title}{Thermodynamic relations for high elastic materials}.
\newblock \emph{\bibinfo{journal}{Transactions of the Faraday Society}}
  \textbf{\bibinfo{volume}{57}}, \bibinfo{pages}{829--838}
  (\bibinfo{year}{1961}).

\bibitem{ortiz_variational_1999}
\bibinfo{author}{Ortiz, M.} \& \bibinfo{author}{Stainier, L.}
\newblock \bibinfo{title}{The variational formulation of viscoplastic
  constitutive updates}.
\newblock \emph{\bibinfo{journal}{Computer Methods in Applied Mechanics and
  Engineering}} \textbf{\bibinfo{volume}{171}}, \bibinfo{pages}{419--444}
  (\bibinfo{year}{1999}).

\bibitem{fancello_variational_2006}
\bibinfo{author}{Fancello, E.}, \bibinfo{author}{Ponthot, J.-P.} \&
  \bibinfo{author}{Stainier, L.}
\newblock \bibinfo{title}{A variational formulation of constitutive models and
  updates in non-linear finite viscoelasticity}.
\newblock \emph{\bibinfo{journal}{International Journal for Numerical Methods
  in Engineering}} \textbf{\bibinfo{volume}{65}}, \bibinfo{pages}{1831--1864}
  (\bibinfo{year}{2006}).

\bibitem{fancello_variational_2008}
\bibinfo{author}{Fancello, E.}, \bibinfo{author}{Ponthot, J.} \&
  \bibinfo{author}{Stainier, L.}
\newblock \bibinfo{title}{A variational framework for nonlinear viscoelastic
  models in finite deformation regime}.
\newblock \emph{\bibinfo{journal}{Journal of Computational and Applied
  Mathematics}} \textbf{\bibinfo{volume}{215}}, \bibinfo{pages}{400--408}
  (\bibinfo{year}{2008}).

\bibitem{mosler_variationally_2010}
\bibinfo{author}{Mosler, J.}
\newblock \bibinfo{title}{Variationally consistent modeling of finite strain
  plasticity theory with non-linear kinematic hardening}.
\newblock \emph{\bibinfo{journal}{Computer Methods in Applied Mechanics and
  Engineering}} \textbf{\bibinfo{volume}{199}}, \bibinfo{pages}{2753--2764}
  (\bibinfo{year}{2010}).

\bibitem{miehe_homogenization_2002}
\bibinfo{author}{Miehe, C.}, \bibinfo{author}{Schotte, J.} \&
  \bibinfo{author}{Lambrecht, M.}
\newblock \bibinfo{title}{Homogenization of inelastic solid materials at finite
  strains based on incremental minimization principles. {Application} to the
  texture analysis of polycrystals}.
\newblock \emph{\bibinfo{journal}{Journal of the Mechanics and Physics of
  Solids}} \textbf{\bibinfo{volume}{50}}, \bibinfo{pages}{2123--2167}
  (\bibinfo{year}{2002}).

\bibitem{platen_nonlinear_2024}
\bibinfo{author}{Platen, J.} \emph{et~al.}
\newblock \bibinfo{title}{A nonlinear finite viscoelastic formulation relative
  to the viscous intermediate configuration applied to plants}.
\newblock \emph{\bibinfo{journal}{International Journal for Numerical Methods
  in Engineering}} \textbf{\bibinfo{volume}{125}}, \bibinfo{pages}{e7483}
  (\bibinfo{year}{2024}).

\bibitem{leuschner_potential-based_2015}
\bibinfo{author}{Leuschner, M.}, \bibinfo{author}{Fritzen, F.},
  \bibinfo{author}{Van~Dommelen, J.} \& \bibinfo{author}{Hoefnagels, J.}
\newblock \bibinfo{title}{Potential-based constitutive models for cohesive
  interfaces: {Theory}, implementation and examples}.
\newblock \emph{\bibinfo{journal}{Composites Part B: Engineering}}
  \textbf{\bibinfo{volume}{68}}, \bibinfo{pages}{38--50}
  (\bibinfo{year}{2015}).

\bibitem{stewart_large_2024}
\bibinfo{author}{Stewart, E.~M.} \& \bibinfo{author}{Anand, L.}
\newblock \bibinfo{title}{A large deformation viscoelasticity theory for
  elastomeric materials and its numerical implementation in the open-source
  finite element program {FEniCSx}}.
\newblock \emph{\bibinfo{journal}{International Journal of Solids and
  Structures}} \textbf{\bibinfo{volume}{303}}, \bibinfo{pages}{113023}
  (\bibinfo{year}{2024}).

\bibitem{li_large-deformation_2022}
\bibinfo{author}{Li, X.}, \bibinfo{author}{Tao, J.}, \bibinfo{author}{Landauer,
  A.~K.}, \bibinfo{author}{Franck, C.} \& \bibinfo{author}{Henann, D.~L.}
\newblock \bibinfo{title}{Large-deformation constitutive modeling of
  viscoelastic foams: {Application} to a closed-cell foam material}.
\newblock \emph{\bibinfo{journal}{Journal of the Mechanics and Physics of
  Solids}} \textbf{\bibinfo{volume}{161}}, \bibinfo{pages}{104807}
  (\bibinfo{year}{2022}).

\bibitem{liu_large_2025}
\bibinfo{author}{Liu, Z.}, \bibinfo{author}{Ortigosa, R.},
  \bibinfo{author}{Gil, A.~J.} \& \bibinfo{author}{Bonet, J.}
\newblock \bibinfo{title}{Large strain constitutive modelling of soft
  compressible and incompressible solids: {Generalised} isotropic and
  anisotropic viscoelasticity}.
\newblock \emph{\bibinfo{journal}{Journal of the Mechanics and Physics of
  Solids}} \textbf{\bibinfo{volume}{203}}, \bibinfo{pages}{106194}
  (\bibinfo{year}{2025}).

\bibitem{d_boor_practical_1978}
\bibinfo{author}{d.~Boor, C.}
\newblock \emph{\bibinfo{title}{A {Practical} {Guide} to {Splines}}}
  (\bibinfo{publisher}{Springer Verlag}, \bibinfo{year}{1978}).

\bibitem{butterfield_computation_1976}
\bibinfo{author}{Butterfield, K.~R.}
\newblock \bibinfo{title}{The {Computation} of all the {Derivatives} of a
  {B}-spline {Basis}}.
\newblock \emph{\bibinfo{journal}{IMA Journal of Applied Mathematics}}
  \textbf{\bibinfo{volume}{17}}, \bibinfo{pages}{15--25}
  (\bibinfo{year}{1976}).

\bibitem{silverman_density_2018}
\bibinfo{author}{Silverman, B.}
\newblock \emph{\bibinfo{title}{Density {Estimation} for {Statistics} and
  {Data} {Analysis}}} (\bibinfo{publisher}{Routledge}, \bibinfo{year}{2018}),
  \bibinfo{edition}{1} edn.

\bibitem{moreno-mateos_learning_2026}
\bibinfo{author}{Moreno-Mateos, M.~A.}, \bibinfo{author}{Wiesheier, S.},
  \bibinfo{author}{Steinmann, P.} \& \bibinfo{author}{Kuhl, E.}
\newblock \bibinfo{title}{Learning ultra-compressible hyperelasticity with
  splines: {Constitutive} asymmetries and non-unique representations}
  (\bibinfo{year}{2026}).
\newblock \bibinfo{note}{ArXiv:2604.14264 [cs]}.

\bibitem{hossain_experimental_2012}
\bibinfo{author}{Hossain, M.}, \bibinfo{author}{Vu, D.~K.} \&
  \bibinfo{author}{Steinmann, P.}
\newblock \bibinfo{title}{Experimental study and numerical modelling of {VHB}
  4910 polymer}.
\newblock \emph{\bibinfo{journal}{Computational Materials Science}}
  \textbf{\bibinfo{volume}{59}}, \bibinfo{pages}{65--74}
  (\bibinfo{year}{2012}).

\end{thebibliography}

\end{document}